\documentclass{article}
\usepackage{natbib,psfig,emulateapj}

\newcommand\nh{\hbox{{$N_{\rm H}\,$}}}
\newcommand\nhgal{\hbox{{$N_{\rm H}^{\rm Gal}\,$}}}

\newcommand\fe{\hbox{{$Z_{\rm Fe}$}}}
\newcommand\solar{\hbox{{$Z_{\sun}$}}}

\newcommand\rosat{{\sl ROSAT} }
\newcommand\asca{{\sl ASCA} }
\newcommand\chandra{{\sl Chandra} }
\newcommand\xmm{{\sl XMM} }

\newcommand\xspec{{\sc xspec} }

\newcommand\ctspers{{ct s$^{-1}$}}
\newcommand\ergcms{{erg cm$^{-2}$ s$^{-1}$}}

\newcommand\kmsmpc{{km s$^{-1}$ Mpc$^{-1}$\thinspace}}

\begin{document}

\title{Iron Gradients in Cooling Flow Galaxies and Groups}
\author{David A. Buote\altaffilmark{1}}
\affil{UCO/Lick Observatory, University of California at Santa Cruz,
Santa Cruz, CA 95064; buote@ucolick.org}
\altaffiltext{1}{Chandra Fellow}

\slugcomment{Accepted for Publication in The Astrophysical Journal}

\begin{abstract}
Previous studies of the Fe abundances in the hot gas of galaxies and
groups have reported conflicting results with most studies finding
very sub-solar Fe abundances that disagree with standard theory.  To
investigate the possible role of Fe abundance gradients on these
measurements we present deprojection analysis of the \rosat PSPC data
of 10 of the brightest cooling flow galaxies and groups.  The PSPC
allows for spatially resolved spectral analysis on a half-arcminute
scale, and interesting constraints on both the temperatures and Fe
abundances are possible because the $\sim 1$ keV temperatures of these
systems are well matched to the bandpass of the PSPC.  In 9 out of 10
systems we find clear evidence that the Fe abundance decreases with
increasing radius: $\fe \approx 1\solar - (\rm several)\solar$ within
the central radial bin ($r\la 10$ kpc) which decreases to $\fe\sim
0.5\solar$ at the largest radii examined ($r\sim 50$-100 kpc).  The Fe
abundances (and temperatures) are consistent with the average values
for these systems that we obtained in our previous analyses of the
\asca data using multitemperature models which confirms that previous
inferences of very sub-solar Fe abundances from \asca arise from the
incorrect assumption of isothermal gas and not the presence of Fe
abundance gradients. We discuss why this ``Fe Bias'' affects much more
seriously the measurements of \fe\, from \asca data than from \rosat
data. We show that the Fe abundance profiles for these galaxies and
groups are consistent with a gas-dynamical model where the gas is
enriched by stellar ejecta and supernovae in the ``solar supernova
proportion'', the stars formed with a Galactic IMF, and the gas is
diluted by mixing with primordial gas at large radii.
\end{abstract}

\section{Introduction}
\label{intro}

One important aspect of the problem of galaxy formation is to
understand the connection between the history of star formation and
the metal enrichment history of the extended hot gaseous component
(e.g., Cavaliere, Giacconi, \& Menci 2000).  Recent studies have
argued that most of the metals produced by star-forming galaxies at
high redshifts exist in the hot gaseous halos of galaxies and
proto-clusters (e.g., Pettini 1999, 2000). By assuming that the hot
gaseous components of nearby rich clusters are a fair representation
of the local universe as a whole, Renzini (1997, 2000) has argued that
the cluster Fe abundances indicate a prompt initial enrichment of the
early universe.

Hence, X-ray observations may be vital for obtaining a complete
picture of the history of star formation of the universe In massive
elliptical galaxies the metals injected by previous generations of
stars do not escape the gravitational field, and thus the hot gas in
these systems records the star formation history via the heating and
enrichment of the hot gas by supernovae (e.g., Ciotti et al 1991;
David, Forman, \& Jones 1991; Loewenstein \& Mathews 1991).  These
systems typically fall into the class of cooling flows (e.g., Fabian
1994).

Unfortunately, reliable constraints on the metal enrichment in such
systems are lacking because the metal abundances in the hot gas in
cooling flow galaxies and groups determined from previous X-ray
studies have yielded conflicting results with most authors finding
very sub-solar Fe abundances (see Buote 2000a and references therein).
These low Fe abundances are generally less than the stellar values
which implies little or no enrichment from Type Ia supernovae and thus
an IMF much flatter than that of the Milky Way (e.g., Arimoto et al
1997; Renzini 1997; Wyse 1997). In contrast, in our recent analyses of
the \asca data accumulated within $r\sim 3\arcmin$-$5\arcmin$ of the
brightest ellipticals and groups we find that the Fe abundances are
approximately solar and that the previous inferences of very sub-solar
Fe abundances using \asca data are the result of a bias arising from
assuming the gas to be isothermal in the presence of significant
temperature gradients (Buote \& Fabian 1998; Buote 1999, 2000a).

To investigate the possible role of Fe abundance gradients on these
measurements we have re-examined the \rosat PSPC data of the brightest
galaxies and groups.  The PSPC allows for spatially resolved spectral
analysis on a half-arcminute scale, and interesting constraints on
both the temperatures and Fe abundances are possible because the $\sim
1$ keV temperatures of these systems are well matched to the bandpass
of the PSPC. Moreover, analysis of the PSPC data is much less subject
to the ``Fe Bias'' \citep{b00a} which plagues analysis of the \asca
data (see \S \ref{bias}).

We selected for re-analysis all of the galaxies and groups possessing
published temperature profiles from \rosat (Forman et al 1993; Ponman
\& Bertram 1993; David et al 1994; Trinchieri et al 1994; Rangarajan
et al 1995; Kim \& Fabbiano 1995; Irwin \& Sarazin 1996; Jones et al
1997; Trinchieri, Fabbiano, \& Kim 1997; Mulchaey \& Zabludoff 1998)
with the exception of the Pegasus I group \citep{tfk} because of the
low S/N. All of these systems have rising temperature profiles and
multitemperature \asca spectra usually attributed to cooling flows.

All but one of these \rosat studies determined the temperatures and
Fe abundances by fitting a single temperature component to the
spectrum in an annulus on the sky. However, each annulus on the sky
contains the projection of emission from larger radii which can
confuse interpretation because of the substantial temperature
gradients. Consequently, we have developed our own deprojection code
to obtain three-dimensional spectral parameters.

The paper is organized as follows. In \S \ref{obs} we present the
observations and discuss the data reduction. The deprojection
procedure is described in \S \ref{deproj} and the results for the
temperature and Fe abundance profiles are given in \S \ref{results};
results for the column densities appear in Buote (2000b, hereafter
PAPER1; 2000c, hereafter PAPER3). Discussion of the results and our
conclusions are presented in \S \ref{disc}.

\section{\rosat Observations and Data Reduction}
\label{obs}

\begin{table*}[t] \footnotesize
\begin{center}
\caption{\rosat Observations\label{tab.obs}}
\begin{tabular}{lccccccc} \tableline\tableline\\[-7pt]
& & \nh & & Date & Exposure & Source Count Rate & Background\\
Name & $z$ & ($10^{20}$ cm$^{-2}$) & Sequence No & (Mo/yr) & (ks) &
(\ctspers) & (\ctspers\thinspace arcmin$^{-2}$)\\ \tableline\\[-7pt]
NGC 507  & 0.01646 & 5.2 & 600254n00 & 8/92 & 5.7/5.3 & 1.1E-1 & 2.8E-4\\
         &         &     & 600254a01 & 1/93 & 15.3/12.6\\
NGC 533  & 0.01810 & 3.2 & 600541n00 & 7/93 & 13.0/11.8 & 9.9E-2 & 2.9E-4\\
NGC 1399 & 0.00483 & 1.3 & 600043n00 & 8/91 & 53.5/18.7 & 2.5E-1 & 4.7E-4\\
NGC 2563 & 0.01630 & 4.3 & 600542n00 & 10/93 & 27.1/20.7 & 2.2E-2 & 3.2E-4\\
         &         &     & 600542a01 & 4/94 & 22.9/20.6\\
NGC 4472 & 0.00290 & 1.7 & 600248n00 & 12/92 & 26.0/22.9 & 2.7E-1 & 9.2E-4\\
NGC 4636 & 0.00365 & 1.8 & 600016n00 & 12/91 & 13.1/6.3 & 3.4E-1 & 1.0E-3\\
NGC 4649 & 0.00471 & 2.2 & 600017n00 & 12/91 & 14.2/9.8 & 2.5E-1 & 1.0E-3\\
NGC 5044 & 0.00898 & 5.0 & 800020n00 & 7/91 & 27.7/19.4 & 5.9E-1 & 5.2E-4\\
NGC 5846 & 0.00610 & 4.2 & 600257n00 & 7/92 & 8.8/7.0 & 1.7E-1 & 7.8E-4\\
         &         &     & 600257a01 & 1/93 & 5.9/3.9\\
HCG 62   & 0.01460 & 2.7 & 800098n00 & 12/91 & 19.6/13.2 & 1.5E-1 &
8.7E-4 \\ \tableline \\[-35pt]
\end{tabular}
\tablecomments{Redshifts are taken from NED. Galactic Hydrogen column
densities (\nh) are taken from \citet{dl} using the HEASARC w3nh
tool. The Exposure column lists first the raw exposure time and second
the final filtered exposure (see \S \ref{obs}).  The source count rate
is computed within the central $R=1\arcmin$. Both the source and
background rates are computed over 0.2-2.2 keV. For systems with
multiple observations the source and background rates are computed
from the total data.}
\end{center}
\end{table*}

We obtained {\sl ROSAT} PSPC data from the public data archive
maintained by the High Energy Astrophysics Science Archive Research
Center (HEASARC). The properties of the observations are listed in
Table \ref{tab.obs}. Except where noted below, these data were reduced
using the standard {\sc FTOOLS} (v4.2) software according to the
procedures described in the OGIP Memo OGIP/94-010 (``ROSAT data
analysis using xselect and ftools''), the WWW pages of the {\sl ROSAT}
Guest Observer Facility (GOF) (see
http://heasarc.gsfc.nasa.gov/docs/rosat), and \citet{pluc}.

The events files of each observation were cleaned of after-pulse
signals by removing all events following within 0.35 ms of a
precursor. To minimize the particle background contribution only
events with Master Veto Rate less than 170 \ctspers\thinspace were
selected \citep{pluc}.  We corrected the Pulse Invariant (PI) bins of
each data set for spatial and long-term temporal gain variations using
the most up-to-date calibration files. For the {\sc ftool} {\sc
pcecor}, which corrects for the variation in the linearity of the PSPC
response, we used the in-flight calibration data for the correction.

From visual inspection of the light curve of an observation we
identified and removed time intervals of significant enhancements in
the count rate in order to obtain a flat distribution of count rate
versus time. Such short-term enhancements are typically the result of
scattered light from the Sun, auroral X-rays, and enhanced charged
particle precipitation \citep{snow2}. The raw and filtered exposure
times are listed in Table \ref{tab.obs}.

The particle background spectrum for an observation was obtained by
following the instructions in \citet{pluc}. We developed our own
software to perform this task because we identified serious errors in
the {\sc ftools} implementation of the Plucinsky et al procedure
(i.e., program {\sc pcparpha}). For our data sets the particle
background rate is alway much less than that of the diffuse
background.

For each observation we obtained a background spectrum from
source-free regions far away from the center of the field (typically
distances of $45\arcmin$-$50\arcmin$). By using a local background
estimate residual contamination from solar X-rays and any other
long-term background enhancements \citep{snow2} are fully accounted
for in the ensuing spectral analysis. However, the extended emission
of the galaxies and groups do contribute at some level to the flux
even at these large radii.

To assess the contribution of galaxy and group emission within the
background regions we fit the background spectrum with a model after
subtracting the particle background. We represent the cosmic X-ray
background by a power law and the Galactic emission by two thermal
components following \citet{cfg}. To account for emission from the
galaxy or group we include another thermal component with variable
temperature.  Each component is modified by the Galactic hydrogen
column listed in Table \ref{tab.obs}.

This composite model provides a good fit to the background spectrum
for each system. In every case the additional thermal component is
required with temperatures ranging from 0.3-1.6 keV. In 8 out of 10
cases this additional component contributes significantly only to
energies above $\sim 0.5$ keV which is consistent with emission from
the galaxy or group in question. For these systems we subtract this
temperature component from the background when performing spectral
analysis (see \S \ref{bkg}), and the background rates listed in Table
\ref{tab.obs} reflect this subtraction. The effect of excluding this
extra component in the background on the source spectral parameters is
only significant for a few systems, and is noticeable only in the
outermost one or two radial bins (see \S \ref{caveats}).

For NGC 5044 and 5846 we obtained $T\sim 0.3$ keV for the extra
thermal component and found that it contributed very significantly to
the emission in both the soft and hard energy channels. In these cases
the extra component probably represents a combination of emission from
long-term enhancements in the background \citep{snow2} with any
residual emission from the groups. As a result, we did not subtract
out the extra thermal component from the background of these systems. 

For NGC 4472 we confirm the finding by \citet{forman} that a
background spectrum taken from regions $\sim 40\arcmin$-$50\arcmin$
from the field center tends to over-subtract energies below $\sim 0.3$
keV probably because of contamination from unresolved point
sources. To compensate for this we took the background from regions
just inside the inner ring of the PSPC where the PSF is much smaller
which allows us to avoid point sources more effectively. We confine
the region to the north of the center of NGC 4472 where the galaxy
emission is lower \citep{is}.

\section{Deprojection Method}
\label{deproj}

The method we use to deproject the X-ray data is the well-established
technique pioneered by \citet{deproj}. This method is non-parametric
in that no functional form is assumed for the spatial distribution of
the X-ray emission or for any of the associated spectral quantities
(temperature, abundances etc.). After first assuming a specific
geometry one begins by determining the emission in the bounding
annulus and then works inwards by subtracting off the contributions
from the outer annuli. For the case of spherical symmetry the volume
emission density is related to the surface brightness according to
the simple geometric formula given by Kriss, Cioffi, \& Canizares
(1983).

The assumption of hydrostatic equilibrium is usually incorporated into
this deprojection procedure in order to obtain the mass and mass
deposition rate as well as the spectral parameters (Fabian et al 1981;
Sarazin 1986; Arnaud 1988). In the interest of generality we do not
make this additional assumption. Hence, the principal assumption in
our version of the deprojection procedure is that of spherical
symmetry which means that our derived spectral parameters should be
considered spherically averaged quantities for systems that have
significant ellipticity.

A thorough, up-to-date discussion of this deprojection method is given
by \citet{deproj_new}, and we refer the reader to that paper for the
relevant equations. Since McLaughlin is interested in the globular
cluster distribution in M87 he formulates the deprojection algorithm
in terms of the number of globular clusters,
$\mathcal{N}(R_{i-1},R_i)$, located between radii $R_{i-1}$ and $R_i$
on the sky. To express his equations in terms of quantities relevant
for X-ray analysis we simply associate $\mathcal{N}(R_{i-1},R_i)$ with
the X-ray flux, $F_{\rm x}(R_{i-1},R_i)$, in \ergcms.

We have developed our own code to implement the deprojection algorithm
and have verified it using synthetic \xmm data obtained by using the
{\sc quicksim} software \citep{quicksim}. Some aspects of the
deprojection analysis require special mention which we now address.

\subsection{Edge Effect}
\label{edge}

The deprojection algorithm assumes there is no source emission outside
of the bounding annulus, $R_m$. This is generally not the case, and if
the exterior emission is not accounted for then the volume emission
density of the outermost annuli will be overestimated.  \citet{nb}
provide an analytic correction factor for the flux of the bounding
annulus in the limit of annuli with zero width. Since, however,
the width of the bounding annulus tends to be large because of
decreasing S/N with increasing radius, and since any exterior emission
also projects into interior annuli, we follow \citet{deproj_new} and
compute a separate correction factor, $f(R_{i-1},R_i)$, for each
annulus arising from emission exterior to $R_m$.

The emissivity of this exterior emission projected into an annulus
$(R_{i-1},R_i)$ is taken to be proportional to $f(R_{i-1},R_i)F_{\rm
x}(R_{m-1},R_m)$; i.e. the spectral shape of the emission exterior to
$R_m$ is assumed to be the same as that for the outermost
annulus. Moreover, to compute $f(R_{i-1},R_i)$ a model for the spatial
profile for the exterior emission is required.  We assume that the
volume X-ray emission density is a power law outside of $R_m$ which is
a good approximation for galaxies and groups since the X-ray emission
at large radii is generally well described by a $\beta$ model
$(r^{-6\beta})$ with $\beta\approx 0.5$-$0.7$ (e.g., Mulchaey \&
Zabludoff 1998).  Our results are not very sensitive to these
assumptions because our chosen bounding annuli are quite wide so that
$f\ll 1$ for $\beta$ over this range.

The formulae for $f(R_{i-1},R_i)$ for the cases $\beta=1/2$ and
$\beta=2/3$ are given by equations A7 and A8 in
\citet{deproj_new}. All results presented in \S \ref{spec} are for the
case $\beta=2/3$ since, as mentioned above, the results are quite
insensitive to this choice.  Both the numerator and denominator of
McLaughlin's equation A8 evaluate to zero for $R_{i-1}=0$ which is
problematical for numerical computations. Fortunately, the limit is
well-behaved, and in Appendix \ref{appendix} we give the limiting
equation.

\subsection{Error Estimation}
\label{error}

The deprojected X-ray emission in any annulus interior to the bounding
annulus depends on the previous results obtained for adjacent exterior
annuli. Since the deprojection procedure introduces correlations
between annuli it is natural to employ Monte Carlo simulations to
estimate the statistical uncertainties on the spectral parameters.
Hence, after we have obtained best-fitting models for all desired
annuli on the sky, we simulate a new data set for each annulus by
using these models as templates. Then we deproject these simulated
data and obtain parameters in the same manner as done for the actual
data.

For each observation we determine the errors on each parameter from
100 Monte Carlo simulations. In most cases we found that the simulated
parameters are reasonably symmetrically distributed about a central
peak. Consequently, we define confidence limits for a parameter with
respect to the median value from the 100 simulations. For example,
after sorting the 100 values for a given parameter into ascending
order we define the 68\% confidence limits to be given by the 16th and
84th values; the 90\% limits are given by the 5th and 95th values,
etc.

This scheme for defining confidence limits has its limitations. In
some cases when the parameter has a large error the simulated
distribution is very flat so that the median is not obviously the best
reference point from which to define confidence limits. In such
instances the 68\% confidence limits defined as above do not always
enclose the best-fitting value. However, we do find that in all cases
the best-fitting values are enclosed with the $\sim 90\%$ confidence
limits indicating that our definitions are not very unreasonable, and
thus we apply the above definitions of confidence levels in every case
for consistency and convenience.

\subsection{Radial Parameter Fluctuations}
\label{fluc}

It is well known from optical studies that the deprojection of the
luminosity distribution of an elliptical galaxy yields a jagged
profile where the departures from a smooth profile are of the same
magnitude or larger than the statistical errors on the photometry
(e.g., Binney, Davies, \& Illingworth 1990). A variety of factors
contribute to the high-frequency noise responsible for the jagged
profile such as the incomplete removal of point sources. An analogous
situation occurs for the deprojection of X-ray data.

In actuality the situation is worse for the X-ray case because one
desires the profiles of several spectral parameters in addition to the
luminosity, but the statistical noise is much greater while the
ability to remove contaminating sources is usually much worse than in
the optical. As a result, the radial fluctuations in parameter values
can sometimes be sufficiently large so that the values get sent off to
obscure regions of parameter space never to return. Such
fluctuations do not agree with the parameters obtained from 
fitting the X-ray spectra without deprojection.

Our preferred means to eliminate such fluctuations is to increase the
S/N by increasing the widths of the annuli. This method is very
effective and has the great advantage that no theoretical prejudice is
forced onto the data (other than the desire for the 2D and 3D profiles
to be qualitatively similar). Although doing this removes most of the
serious parameter fluctuations in our \rosat observations, in many
cases there is still one annulus with divergent parameter values.

Hence, to keep any remaining fluctuations in check we smooth the
derived parameter distributions in a manner related to linear
regularization (e.g., Press et al 1992). The problem with linear
regularization is that one restricts the point-to-point variation in
the parameters with a pre-conceived model. For example, \citet{fp}
assume the smooth profiles of the temperature and abundances to be
linear functions of logarithmic radius. (The amount of smoothing
applied also depends on an additional weighting factor -- see
Finoguenov \& Ponman). However, as we show in \S \ref{fp} these
choices usually lead to very biased parameter profiles.

Our method of regularizing the parameters is not standard since we do
not add an extra term to the $\chi^2$ equation because it cannot be
done in the current version of \xspec \citep{xspec}. Instead we
perform an {\it ex post facto} regularization by simply restricting the
available range for the temperatures and Fe abundances at a given
radius according to the desired amount of smoothing. This method is
simpler to implement and, we believe, allows more transparent control
over the smoothing process.

To insure that we do not over-smooth our parameter profiles we do the
following: (1) We always compare the results obtained from analysis
with and without deprojection and require that the two cases do not
differ qualitatively. (2) We only regularize the temperatures and
Fe abundances and always vary the amount of smoothing for each
observation until we are satisfied that significant bias is not
introduced. In most cases we found that restricting the absolute
values of the radial logarithmic derivatives in the temperature and
Fe abundance to be less than 1 and 1.5 respectively worked well.

\subsection{$\chi^2$ Issues}
\label{chi}

All spectral fitting was performed with the software package \xspec
\citep{xspec} using the ${\chi^2}$ method implemented in its standard
form. The weights for the ${\chi^2}$ method are computed in each PI
bin assuming gaussian statistics. To insure that the weights are valid
in this assumption we regrouped the PI bins for each source spectrum
so that each group has at least 30 counts. The extraction of spectra
is discussed in the following section.

A possible concern with the interpretation of $\chi^2$ for
goodness-of-fit is that the deprojection procedure causes the emission
at inner radii to depend on that from larger radii. However, the model
representing the projected emission within a given annulus is taken to
be exact; i.e. statistical errors from the projected emission model
are not included in the deprojection method, and therefore the
statistical errors in each PI bin remain those of the 2D annulus which
are gaussian. We have verified the validity of using $\chi^2$ for
measuring goodness-of-fit under the standard assumption of gaussian
statistics by comparing to results obtained from 2D analysis without
deprojection: in all cases investigated we find that when a large
improvement in $\chi^2$ is obtained from the deprojection analysis a
qualitatively consistent large improvement is also obtained with the
2D analysis.

\section{Spectral Analysis}
\label{spec}

\subsection{Preliminaries}
\label{pre}

\subsubsection{Extraction of Annular Spectra}

For each system we extracted spectra in concentric circular annuli
located at the X-ray centroid (computed within a $2\arcmin$ radius)
such that for each annulus the width was $\ge 1\arcmin$ and the
background-subtracted counts was larger than some value chosen to
minimize uncertainties on the spectral parameters for each system
while maintaining as many annuli as possible. Data with energies
$\le0.2$ keV were excluded to insure that the PSF was $<1\arcmin$
FWHM. For our on-axis sources $\sim 99\%$ of the PSF at 0.2 keV is
contained within $R=1\arcmin$ \citep{pspcpsf}.

For systems possessing more than one observation (see Table
\ref{tab.obs}) we extracted the source and background spectra
separately from each observation and then added them together. In all
such cases the observations occurred after the October 1991 gain
change so the Redistribution Matrix File (RMF), which specifies the
channel probability distribution for a photon, is the same for all of
them. Since, however, the detector location of the annuli are in
general slightly different for the different observations because of
slight aspect differences, we averaged their respective Auxiliary
Response Files (ARFs) which contain the information on the effective
area as a function of energy and detector location. Any background
sources that were identified by visual examination of the image were
masked out before the extraction.

\subsubsection{Models}

Since the X-ray emission of the galaxies and groups in our sample is
dominated by hot gas, we use coronal plasma models as the basic
component of our spectral models. We use the MEKAL plasma code which
is a modification of the original MEKA code (Mewe, Gronenschild, \&
van den Oord 1985; Kaastra \& Mewe 1993) where the Fe L shell
transitions crucial to the X-ray emission of ellipticals and groups
have been re-calculated \citep{mekal}; the superior performance of the
MEKAL model over the Raymond-Smith code for spectral analysis of
elliptical galaxies we have previously discussed in detail in
\citet{b99}. Because of the limited energy resolution of the PSPC, we
focus on a ``single-phase'' description of the X-ray emission in which
a single temperature component exists at each (three-dimensional)
radius. As we discuss below in \S \ref{bias} we do not expect a
significant ``Fe Bias'' resulting from the single-phase analysis. We
did investigate multiphase models of the hot gas such as a
two-temperature plasma and a constant-pressure cooling flow
\citep{rjcool}, but interesting constraints were not obtained as
explained below in \S \ref{caveats}; see \S 3.1.1 of \cite{b00a} for
further description of such models.

We account for absorption by our Galaxy using the photo-electric
absorption cross sections according to \citet{phabs}. Although
\citet{ab} point out that the He cross section at 0.15 keV is in error
by 13\%, since we analyze $E> 0.2$ keV we find that our fits do not
change when using the \citet{wabs} cross sections which have the
correct He value.  The absorber is modeled as a uniform screen at zero
redshift with solar abundances. The hydrogen column density of the
absorber is generally allowed to be a free parameter to indicate any
additional absorption due to, e.g., intrinsic absorbing material,
calibration errors, etc. We refer to this as the standard absorber
model. (In PAPER3 we discuss more complex absorption models.)

As discussed in PAPER1 and PAPER3 we also find it useful to consider
absorption due to an oxygen edge which we represent by the simple
parameterization, $\exp\left[-\tau(E/E_0)^{-3}\right]$ for $E\ge E_0$,
where $E_0$ is the energy of the edge in the rest frame of the galaxy
or group and $\tau$ is the optical depth; i.e. $\tau$ represents an
absorbing screen located at the source redshift but placed in front of
the source. Partial covering models are discussed in PAPER3.

Finally, data with energies between 0.2 and 2.2 keV were included in
the analysis.

\subsubsection{Meteoritic Solar Abundances}
\label{solar}

\citet{im} have pointed out that the accepted value for the solar Fe
abundance relative to H is approximately $3.24\times 10^{-5}$ by
number. This value is often called the meteoritic value since it was
originally obtained for meteorites, but it also agrees with recent
measurements from the solar photosphere. This is to be contrasted with
the old photospheric value of Fe/H of $4.68\times 10^{-5}$ \citep{ag}
that is widely used. We have decided to use the correct ``meteoritic''
Fe abundance for this and subsequent papers; in \xspec we use the
abundance table of \citet{feld}. However, when comparing to previous
results we shall always take care to consider the factor of 1.44
between the different Fe/H values.

Iron is the only abundance that we allow to be a free parameter. All
other elemental abundances are tied to iron in their solar ratios;
i.e. the values of the other abundances vary with Fe in their fixed
solar ratios.  In a sense we are actually fitting a metallicity, but
since the Fe L-shell lines dominate all other lines in the PSPC
spectrum for the $\sim 1$ keV plasmas of bright galaxies and groups,
the fitted metallicity is almost entirely determined by Fe.  Hence, we
shall always quote our results as Fe abundances.

\subsubsection{Scaling the Background to the Source Position}
\label{bkg}

The observed background spectrum obtained from regions far away from
the source in general covers a different amount of detector area,
requires a different detector response, and if taken from a different
observation can have a different exposure time. Within \xspec the
effect of different areas and exposure times for source and background
are taken into account but not different detector responses. Background
spectra taken near the edge of the PSPC field suffer from vignetting
which significantly reduces the count rate with respect to the source
spectra near the field center. This effect is also energy dependent,
and thus the spectral shape of the background is altered as well.

To properly scale the background spectrum to the source position we
perform the following simple procedure. First, as explained in \S
\ref{obs} we fit a model to the background spectrum. Let us denote the
flux of this model for an energy, $E$, by, $F_E(r_b,A_b,t_b)$, where
$r_b$ represents the detector position, $A_b$ the background area on
the detector, and $t_b$, the exposure time. The corresponding flux in
a given PI channel, $i$, predicted by this model is then,
\begin{displaymath}
F_i(r_b,A_b,t_b) = \sum_E F_E(r_b,A_b,t_b){\rm RMF}_{Ei}(r_b){\rm
ARF}_E(r_b). 
\end{displaymath}
The model, $F_E$, is independent of the response, and thus the
expected background flux in channel, $i$, at the source position is
simply,
\begin{eqnarray} 
F_i(r_s,A_s,t_s) = \hspace{5.7cm} \nonumber \\
\sum_E \left(F_E(r_b,A_b,t_b){A_st_s\over A_bt_b}\right)
{\rm RMF}_{Ei}(r_s){\rm ARF}_E(r_s),  \hspace{0.2cm} \label{eqn.bkg}
\end{eqnarray}
where $r_s$, $A_s$, and $t_s$ are respectively the position, area, and
exposure time for the source.

In the general case one would prefer to use the actual background data
rather than a best-guess model. To do this simply scale the real
background spectrum by the factors $F_i(r_s,A_s,t_s)/
F_i(r_b,A_b,t_b)$ for each PI channel. These ratios are typically very
insensitive to the detailed shape of the input model spectrum.

This scaling procedure was followed for NGC 5044 and 5846. For the
other systems we needed to subtract a thermal component due to the
extended source emission as discussed in \S \ref{tab.obs}. Since for
these systems the background estimate is necessarily defined by a
model we used equation (\ref{eqn.bkg}) for the scaling.

\subsection{Results}
\label{results}

\begin{figure*}[t]
\centerline{\large\bf NGC 507} \vskip 0.1cm
\parbox{0.32\textwidth}{
\centerline{\psfig{figure=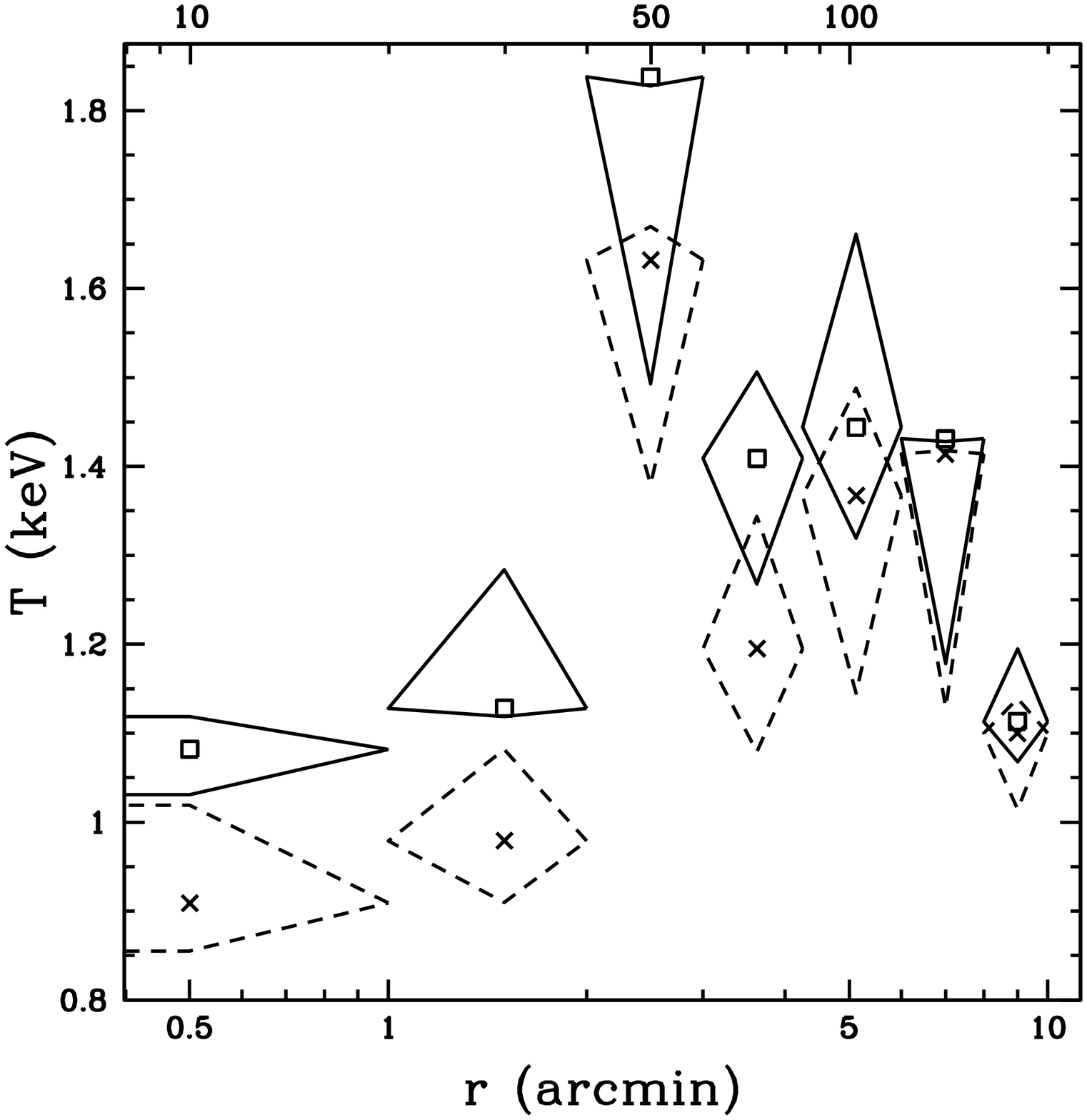,height=0.22\textheight}}
}
\parbox{0.32\textwidth}{
\centerline{\psfig{figure=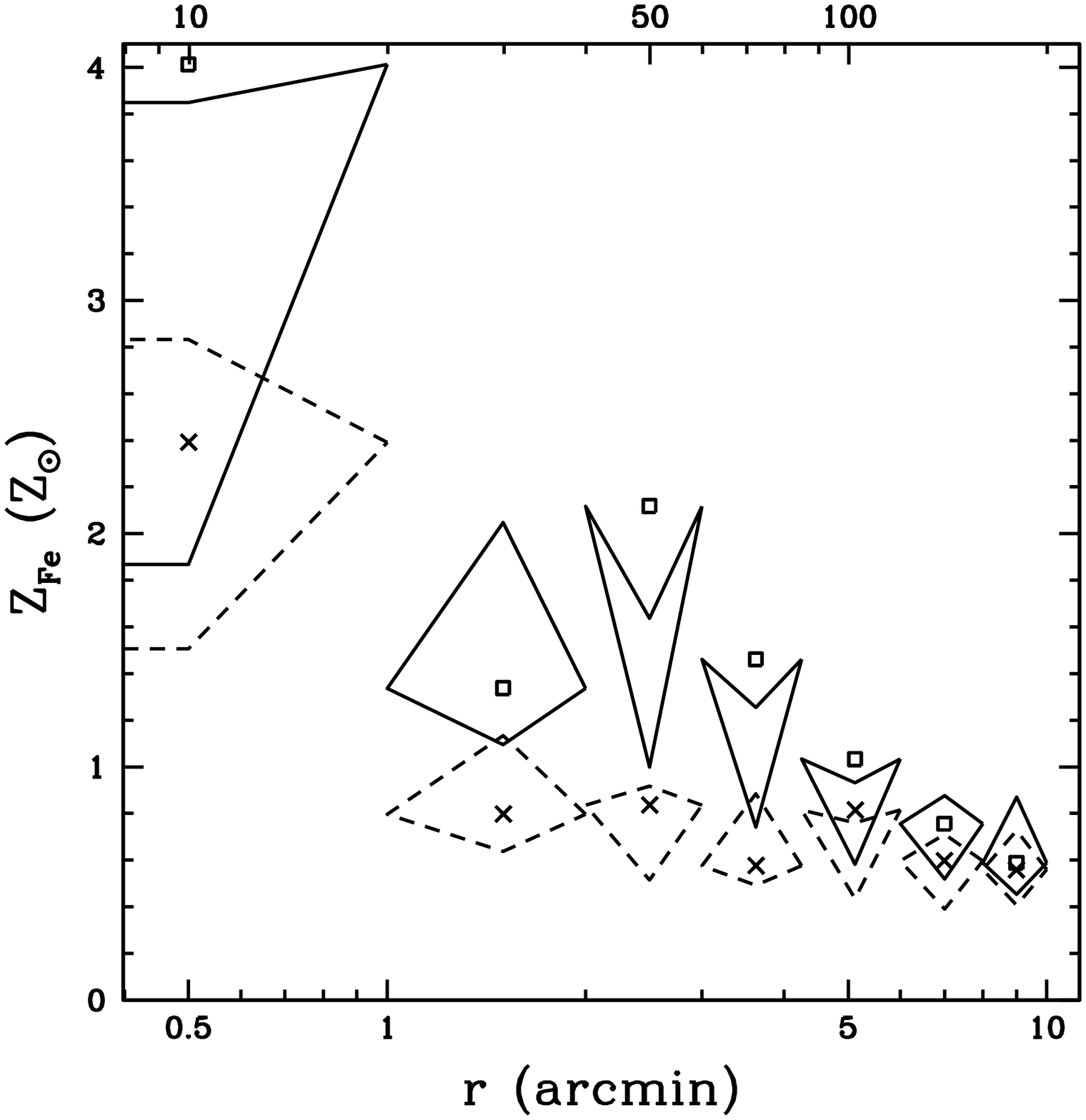,height=0.22\textheight}}
}
\parbox{0.32\textwidth}{
\centerline{\psfig{figure=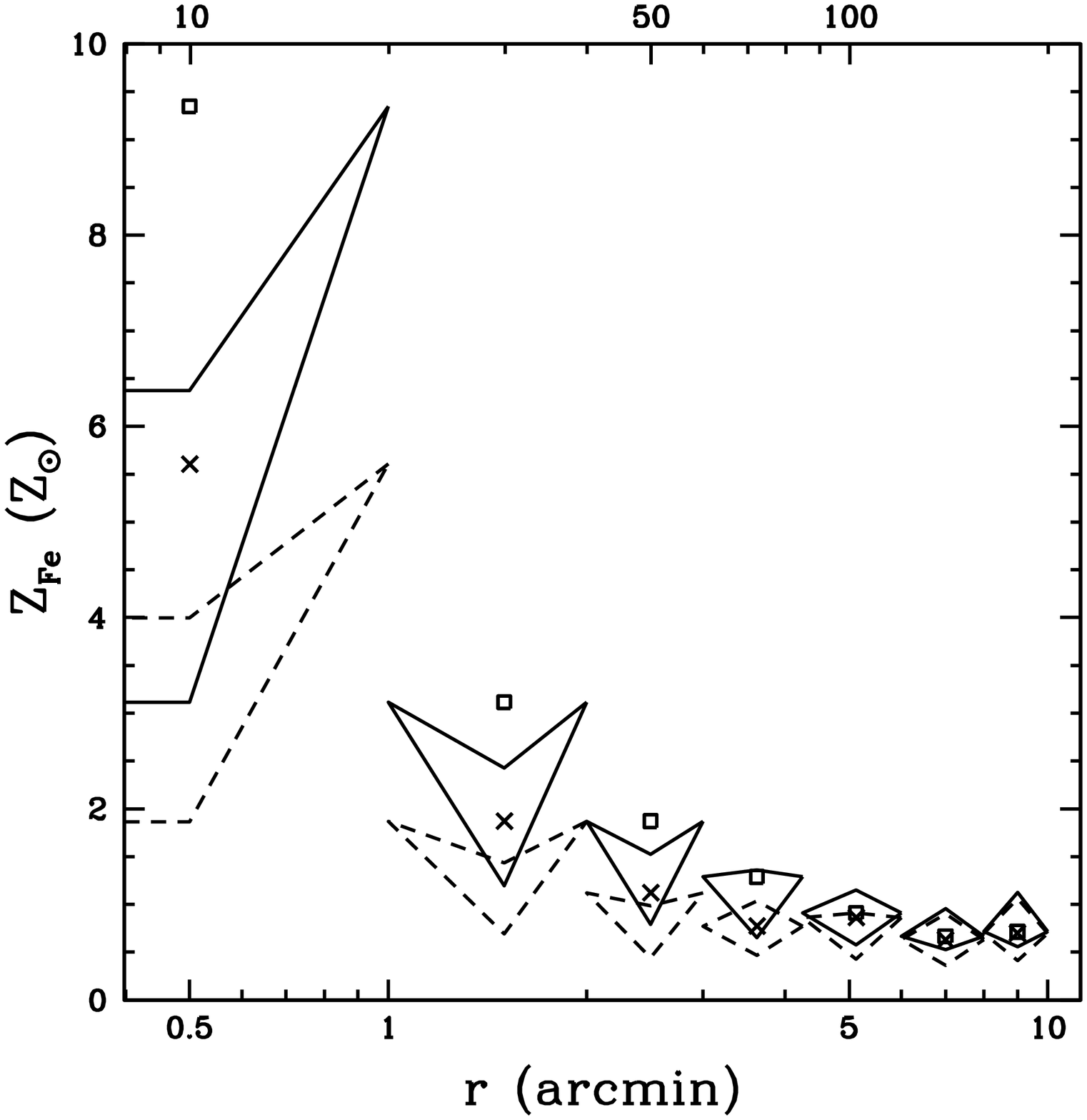,height=0.22\textheight}}
}
\vskip 0.25cm
\centerline{\large\bf NGC 1399} \vskip 0.1cm
\parbox{0.32\textwidth}{
\centerline{\psfig{figure=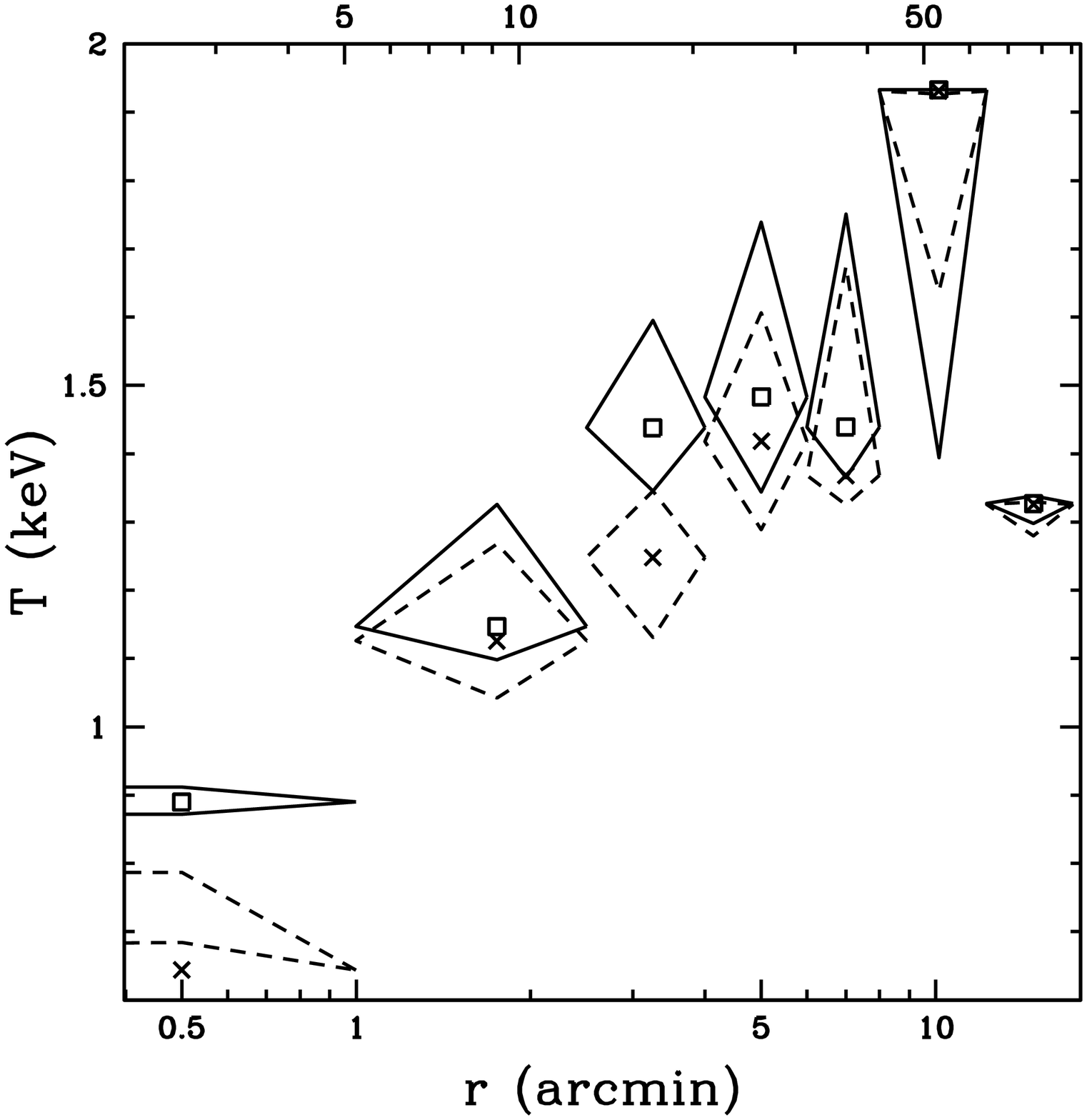,height=0.22\textheight}}
}
\parbox{0.32\textwidth}{
\centerline{\psfig{figure=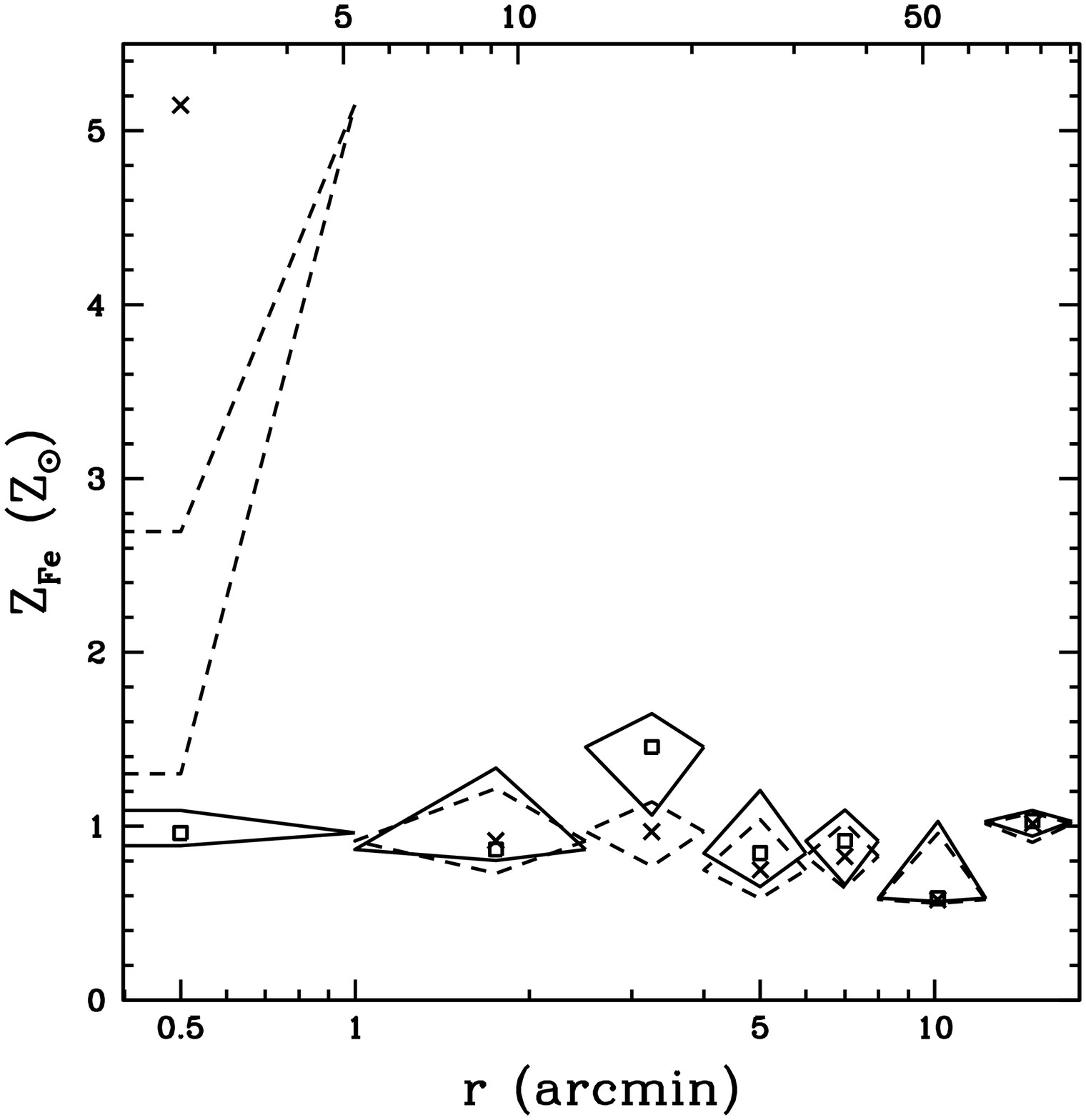,height=0.22\textheight}}
}
\parbox{0.32\textwidth}{
\centerline{\psfig{figure=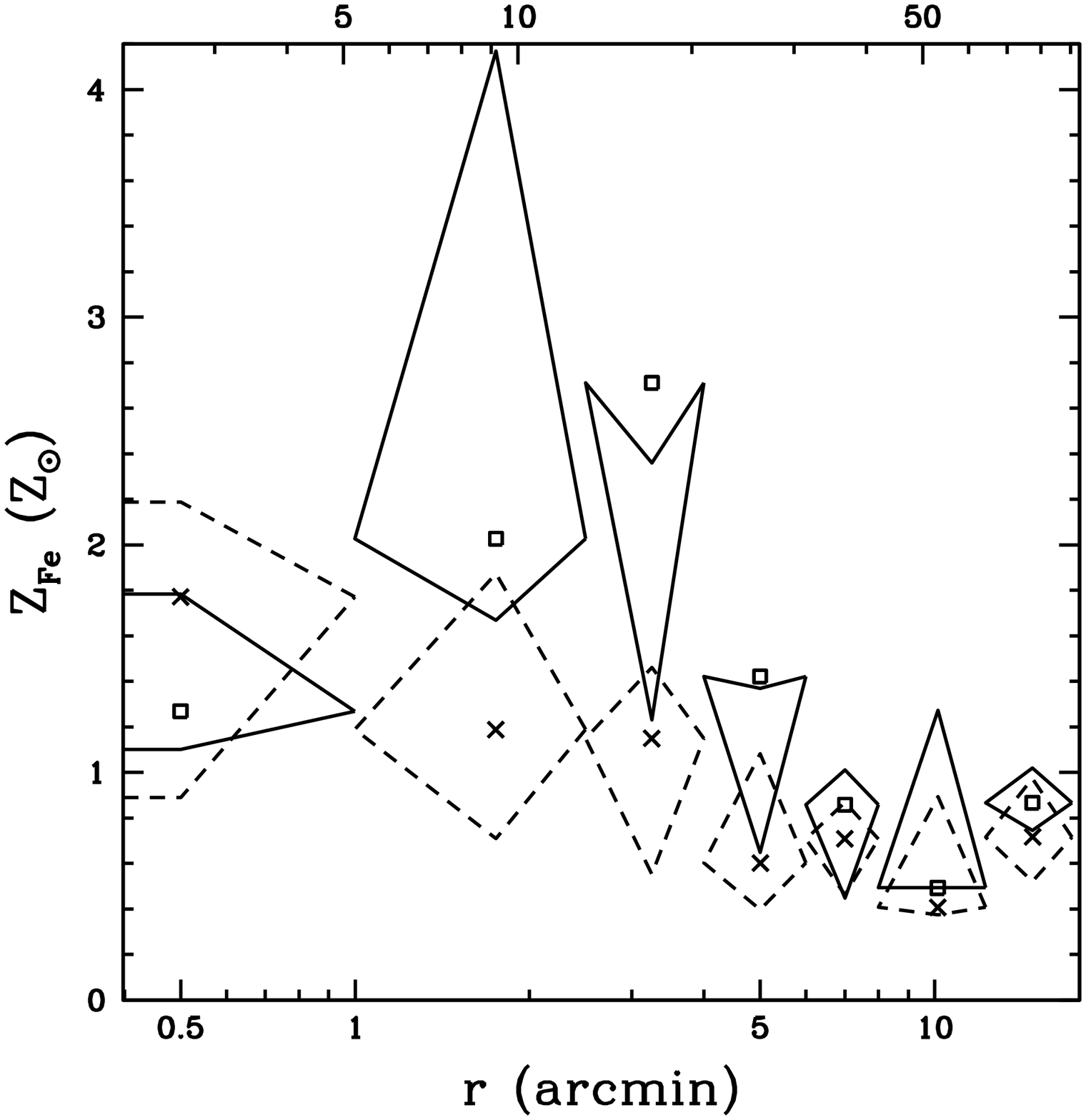,height=0.22\textheight}}
}
\vskip 0.25cm
\centerline{\large\bf NGC 5044} \vskip 0.1cm
\parbox{0.32\textwidth}{
\centerline{\psfig{figure=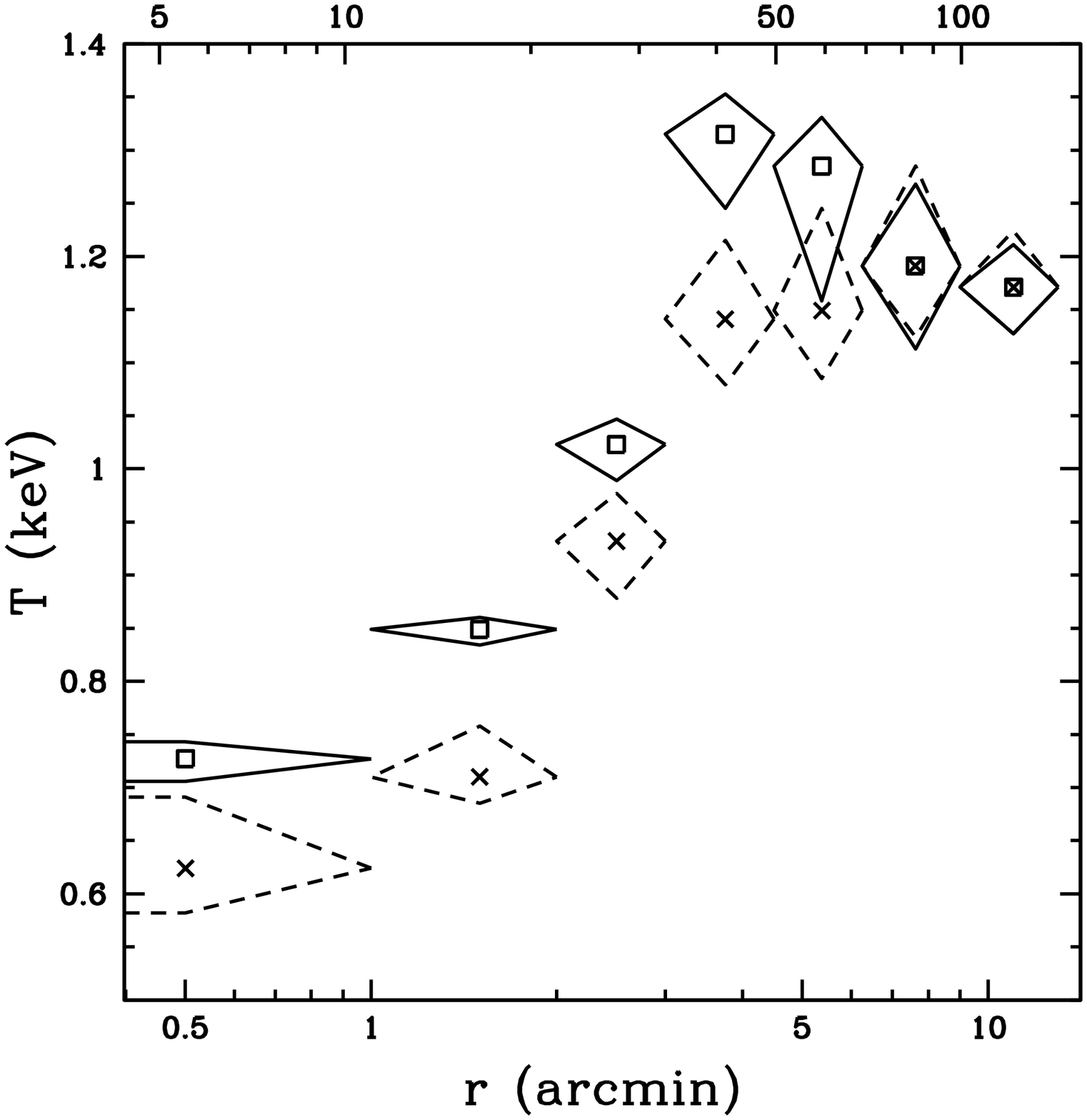,height=0.22\textheight}}
}
\parbox{0.32\textwidth}{
\centerline{\psfig{figure=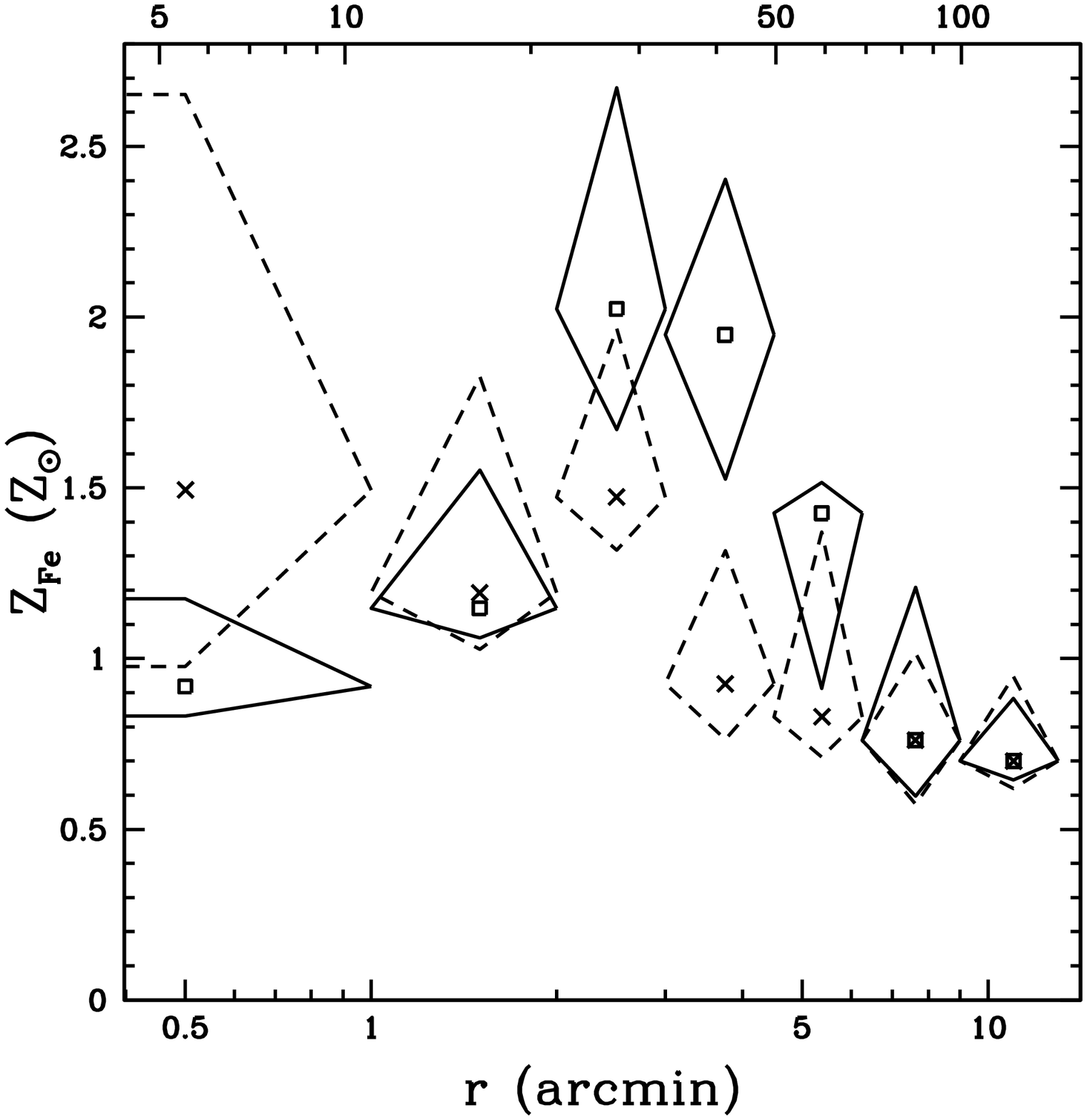,height=0.22\textheight}}
}
\parbox{0.32\textwidth}{
\centerline{\psfig{figure=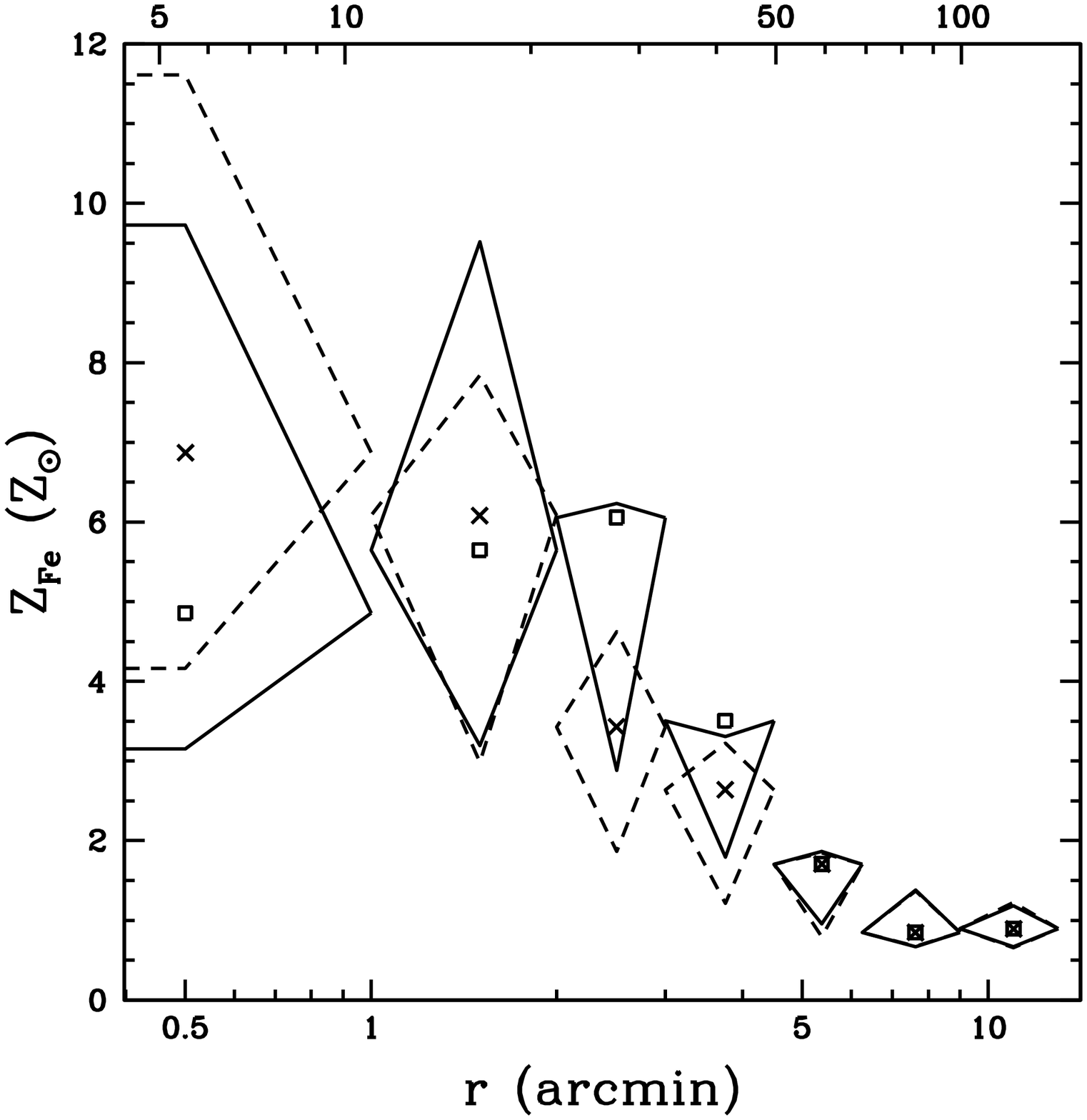,height=0.22\textheight}}
}
\caption{\label{fig.7ann} Results of the deprojection analysis for the
systems where interesting constraints were obtained in 7 annuli. (Left
panels) The temperature profiles for models without an oxygen edge are
denoted by open squares for best fit and solid diamonds for $1\sigma$
error bars; models with an edge are represented by crosses and dashed
diamonds. The column density of the standard absorber is fixed to the
Galactic value in both cases. (Middle panels) The Fe abundance
profiles corresponding to the temperature profiles. (Right panels)
Same as the middle panels except the column density of the standard
absorber is a free parameter. The radial units on the x-axis are
arcminutes on the bottom and kpc on the top obtained from the
redshifts assuming $H_0=70$ \kmsmpc and $\Omega_0=0.3$ except for the
Virgo and Fornax systems for which a distance of 18 Mpc was
assumed. See \S \ref{error} for details on the 68\% confidence error
estimates.}
\end{figure*}

\begin{figure*}[t]
\centerline{\large\bf NGC 2563} \vskip 0.1cm
\parbox{0.32\textwidth}{
\centerline{\psfig{figure=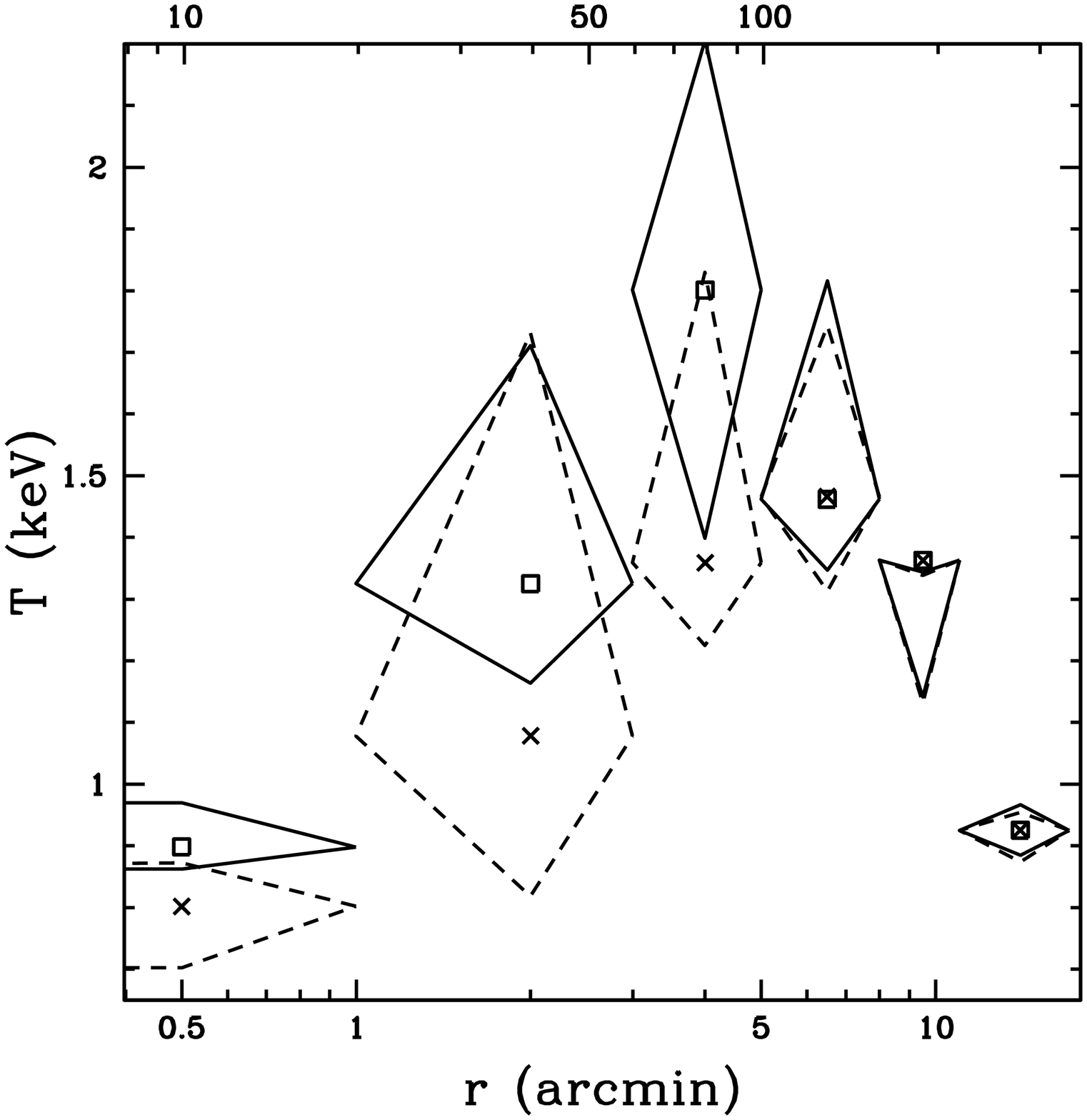,height=0.22\textheight}}
}
\parbox{0.32\textwidth}{
\centerline{\psfig{figure=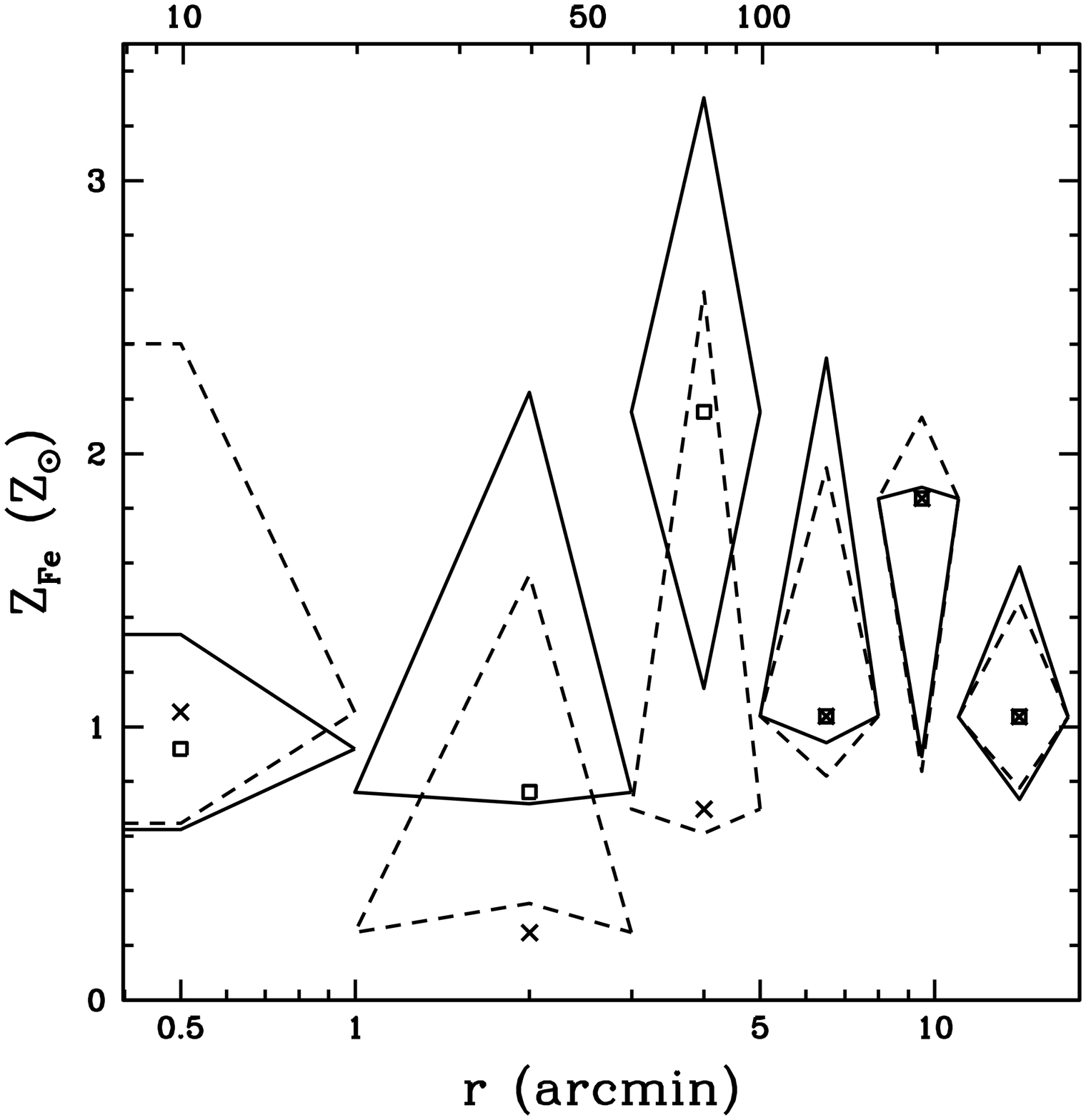,height=0.22\textheight}}
}
\parbox{0.32\textwidth}{
\centerline{\psfig{figure=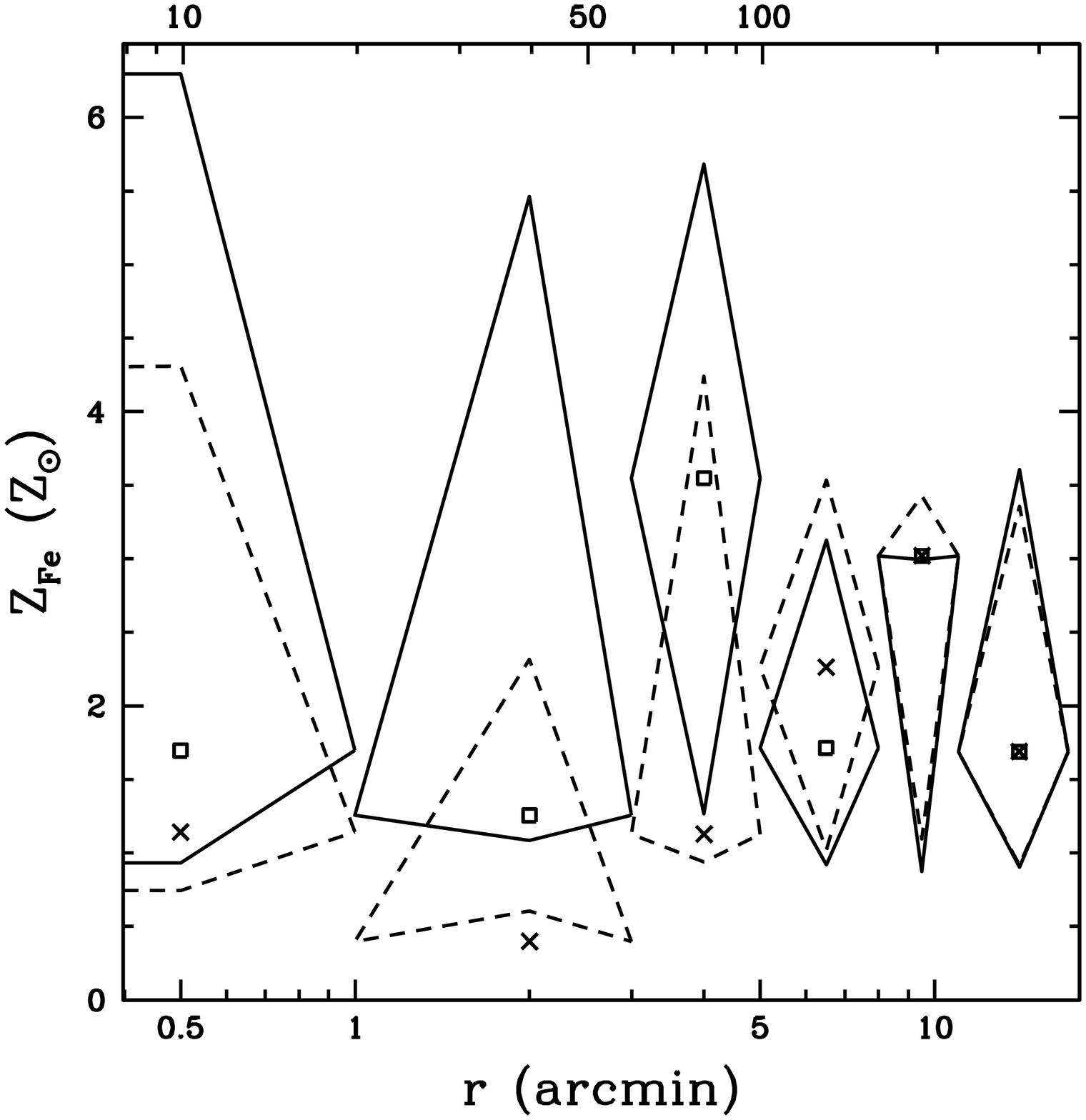,height=0.22\textheight}}
}
\vskip 0.25cm
\centerline{\large\bf NGC 4472} \vskip 0.1cm
\parbox{0.32\textwidth}{
\centerline{\psfig{figure=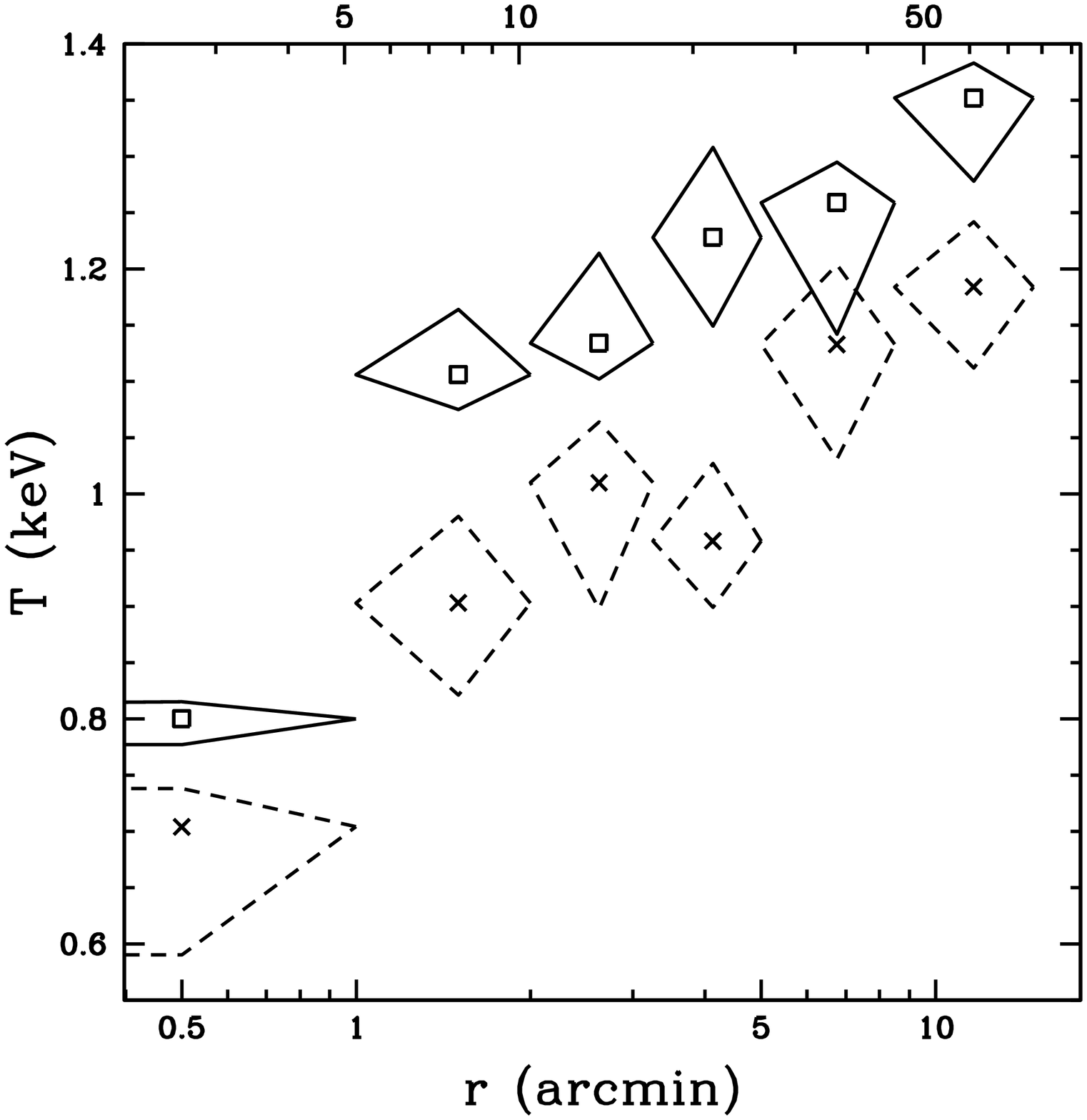,height=0.22\textheight}}
}
\parbox{0.32\textwidth}{
\centerline{\psfig{figure=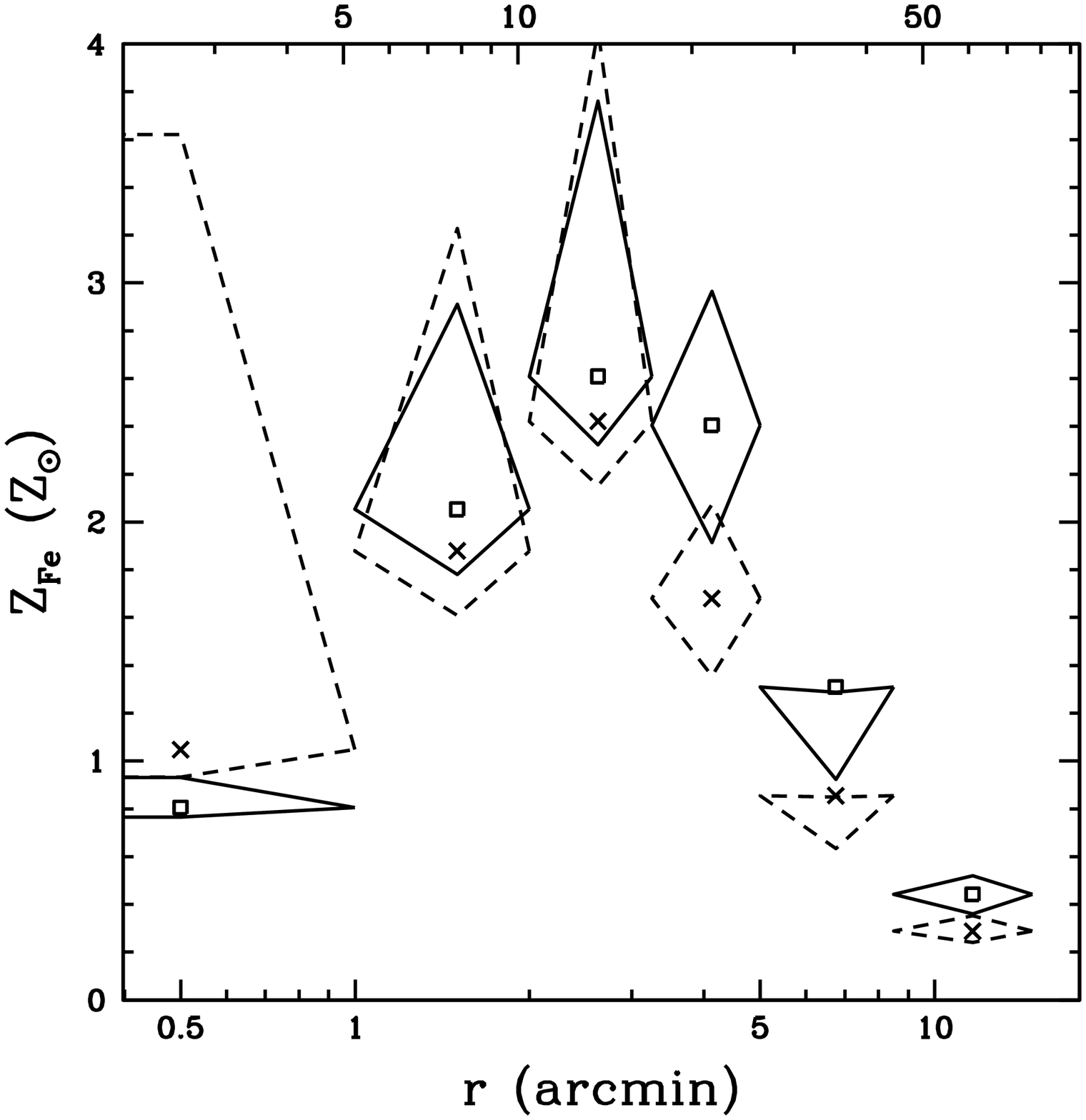,height=0.22\textheight}}
}
\parbox{0.32\textwidth}{
\centerline{\psfig{figure=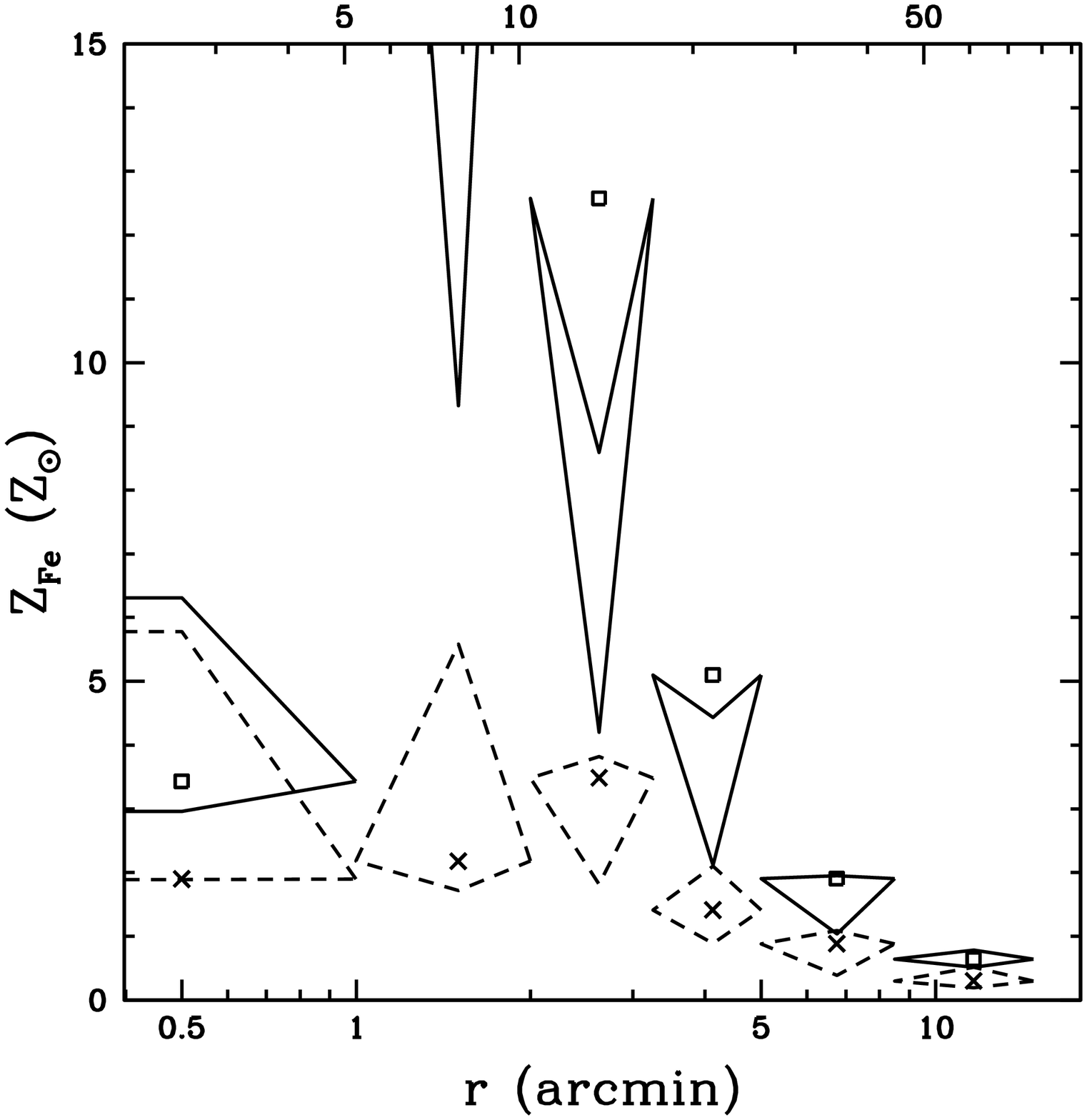,height=0.22\textheight}}
}
\vskip 0.25cm
\centerline{\large\bf NGC 5846} \vskip 0.1cm
\parbox{0.32\textwidth}{
\centerline{\psfig{figure=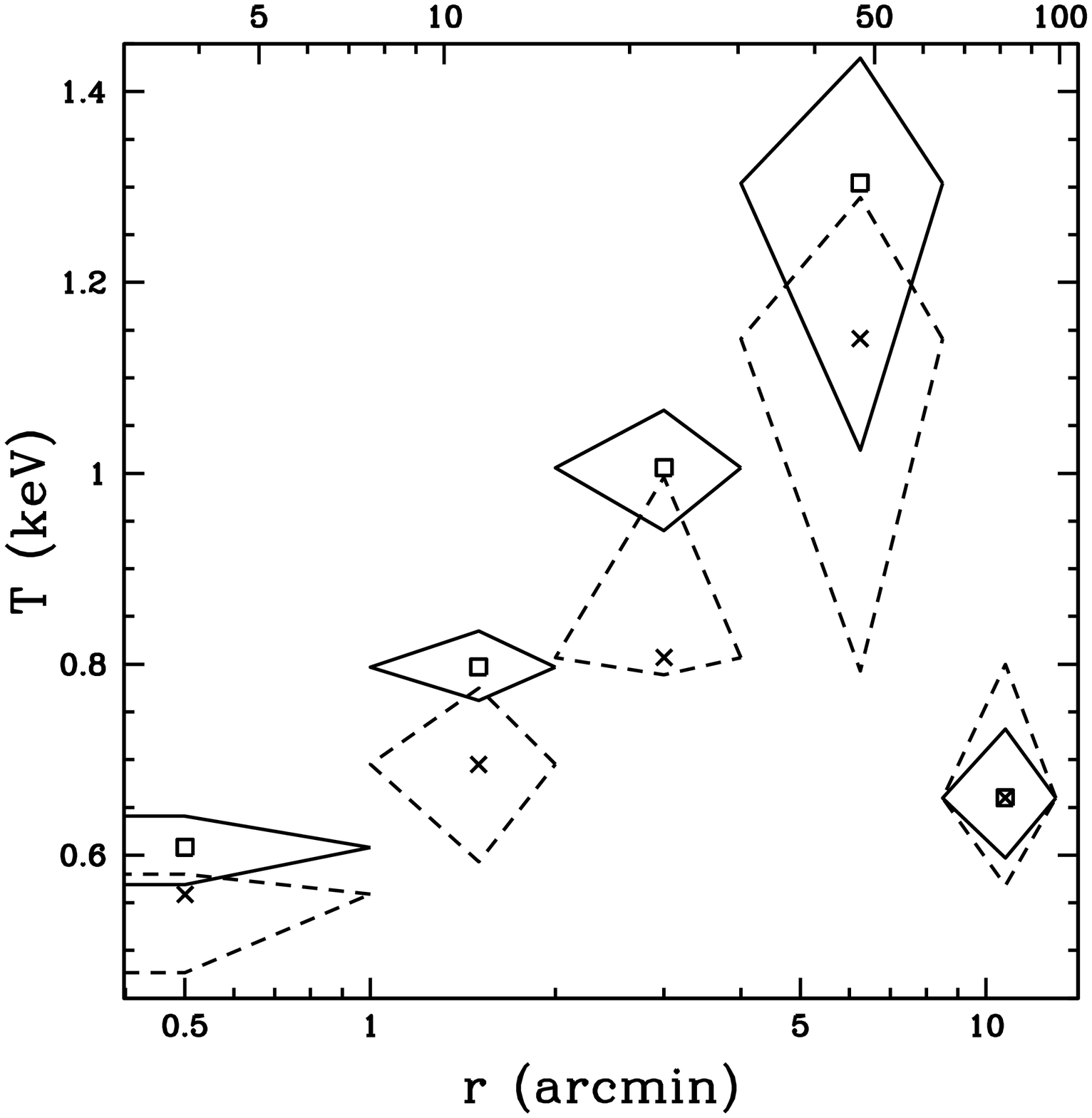,height=0.22\textheight}}
}
\parbox{0.32\textwidth}{
\centerline{\psfig{figure=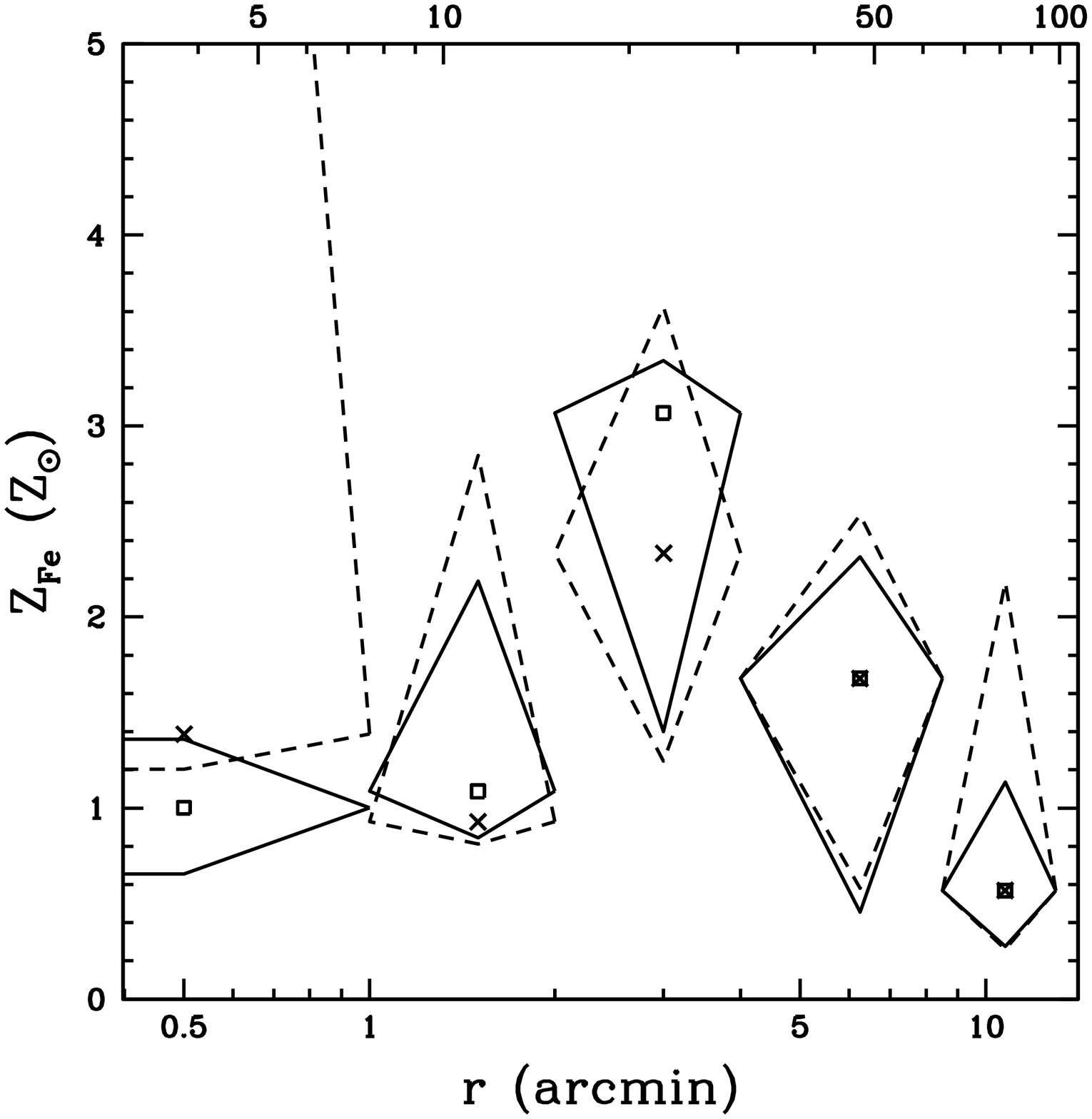,height=0.22\textheight}}
}
\parbox{0.32\textwidth}{
\centerline{\psfig{figure=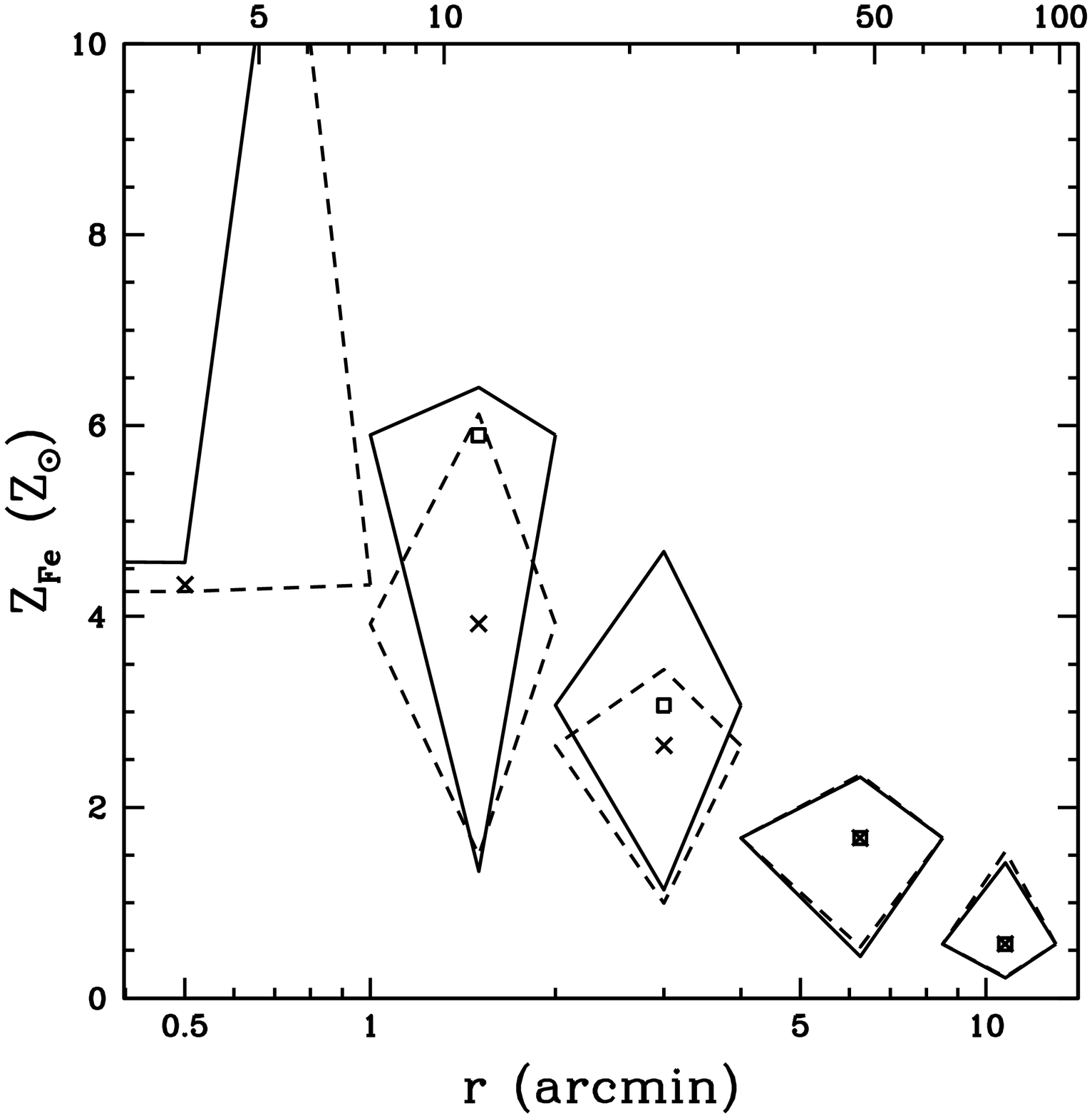,height=0.22\textheight}}}
\caption{\label{fig.6ann} As Figure \ref{fig.7ann} but for systems
with 5 or 6 annuli.}
\end{figure*}

\begin{figure*}[t]
\centerline{\large\bf NGC 533} \vskip 0.1cm
\parbox{0.32\textwidth}{
\centerline{\psfig{figure=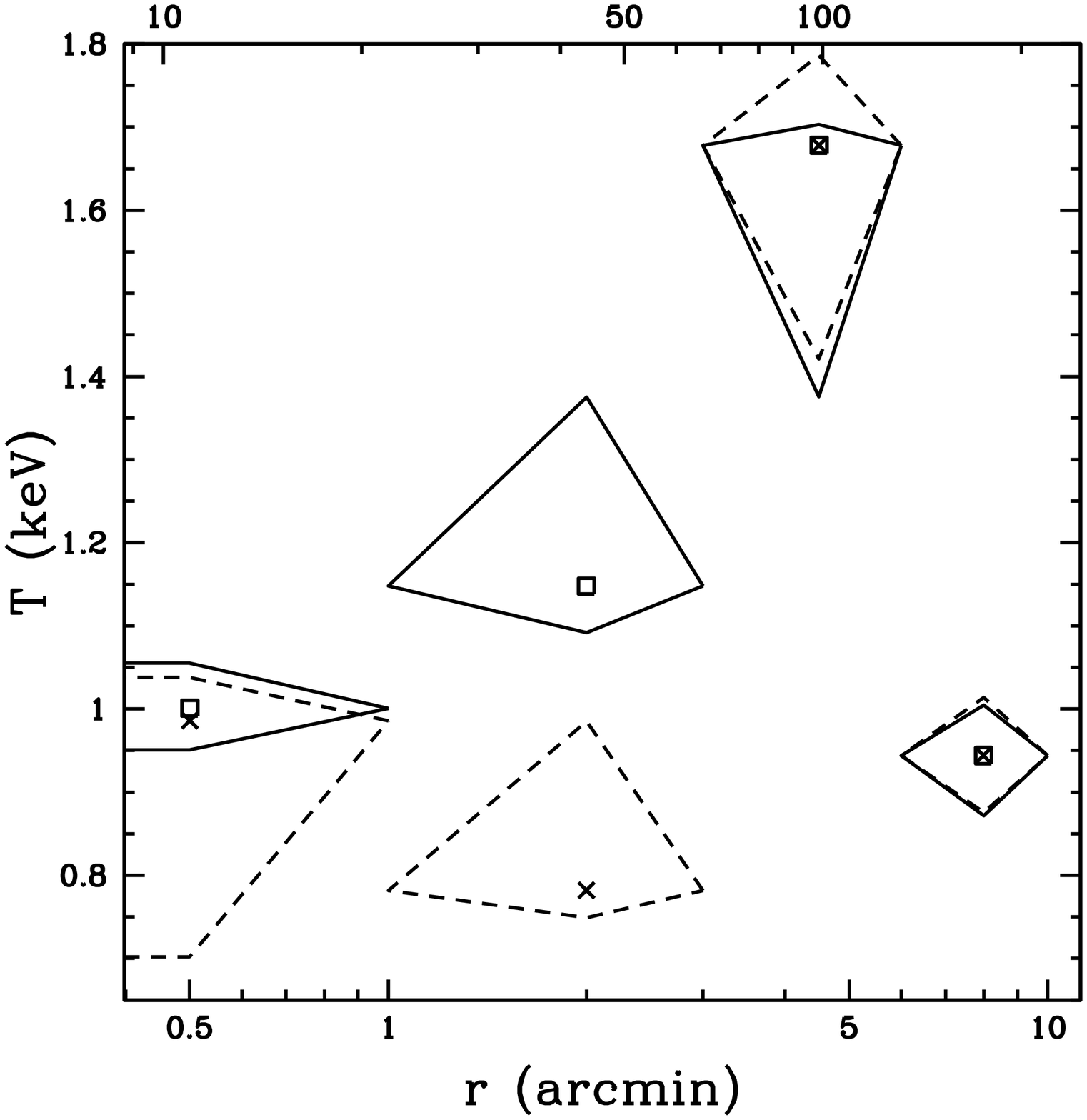,height=0.22\textheight}}
}
\parbox{0.32\textwidth}{
\centerline{\psfig{figure=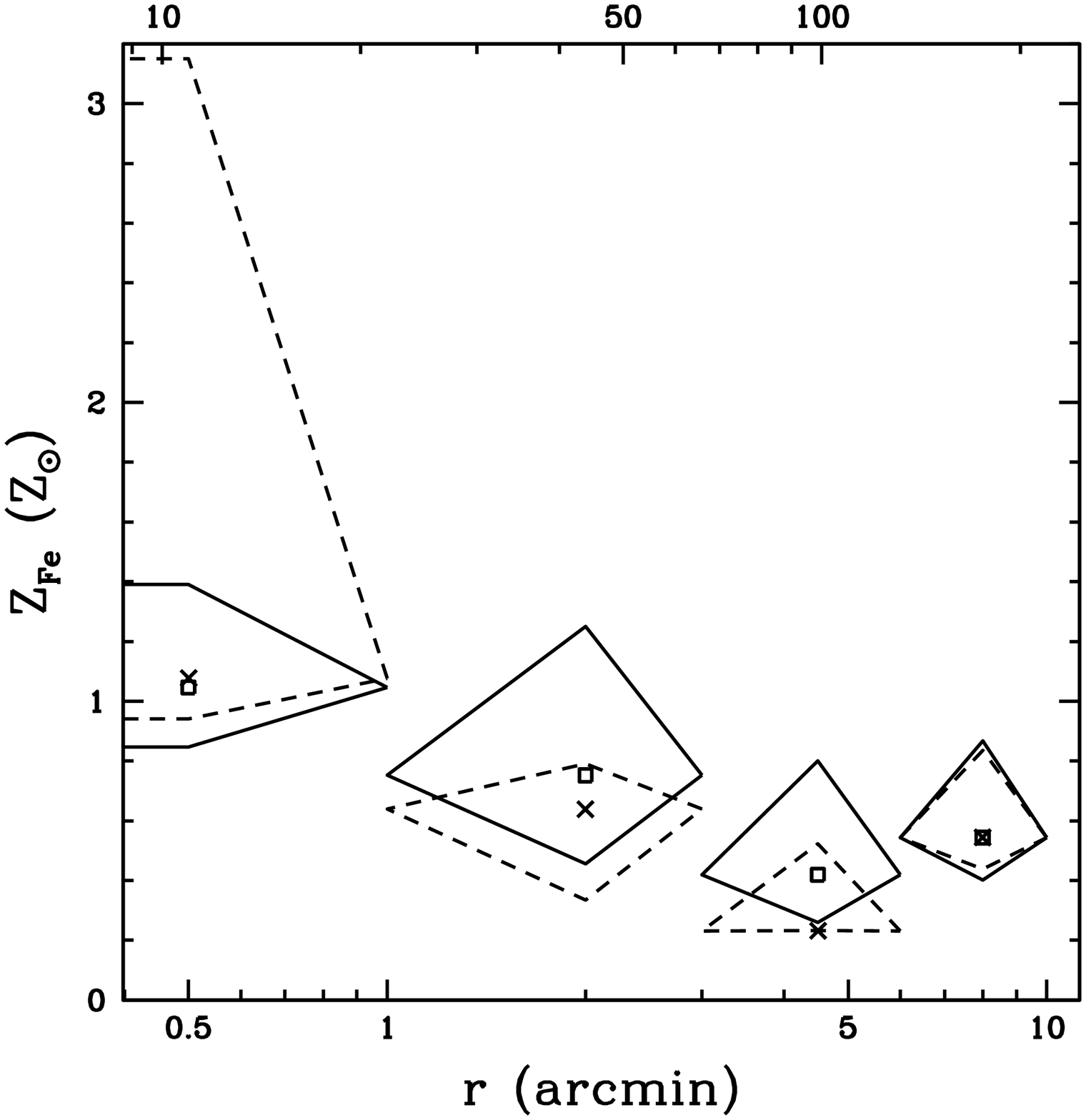,height=0.22\textheight}}
}
\parbox{0.32\textwidth}{
\centerline{\psfig{figure=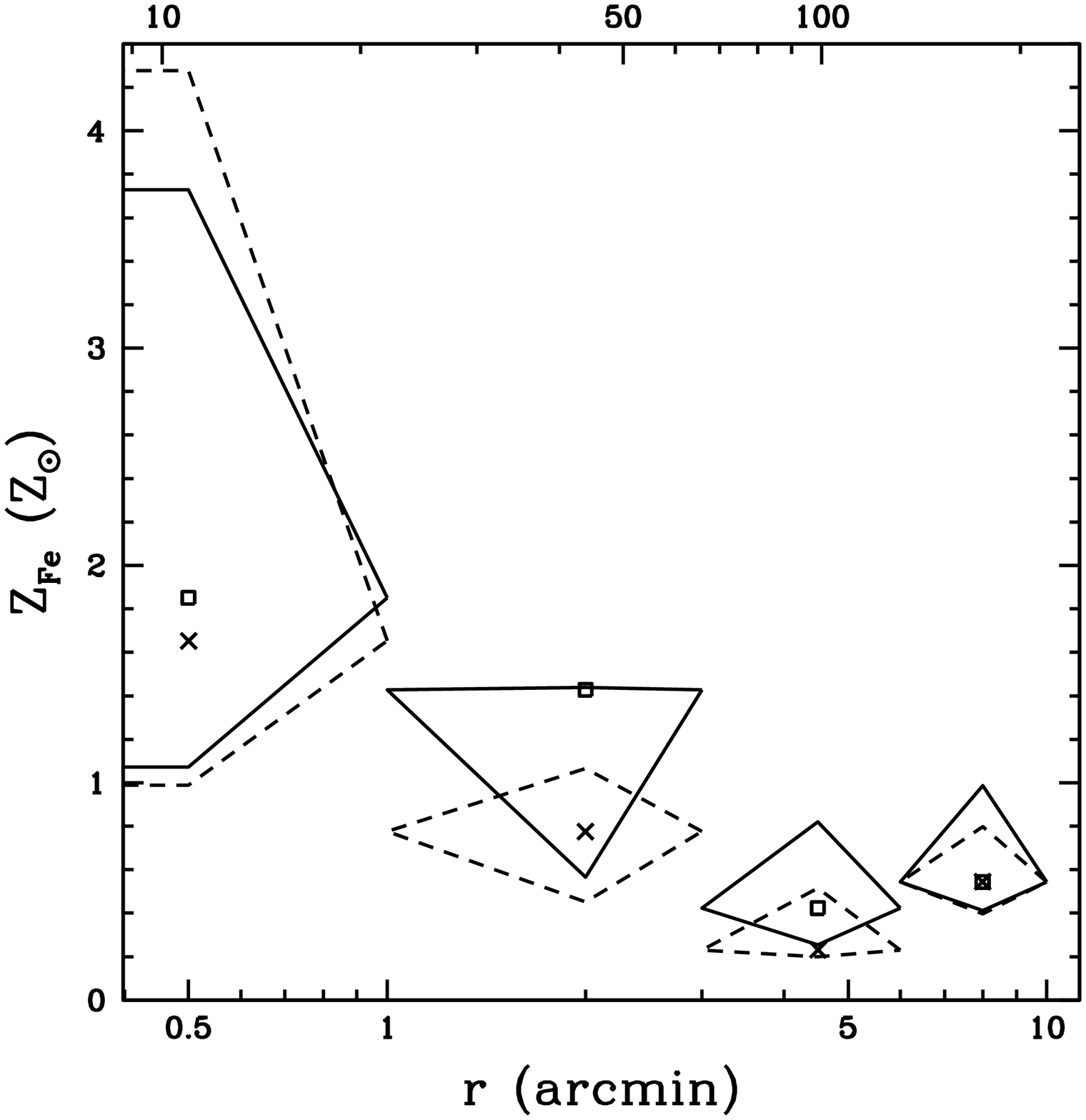,height=0.22\textheight}}}
\vskip 0.25cm
\centerline{\large\bf NGC 4636} \vskip 0.1cm
\parbox{0.32\textwidth}{
\centerline{\psfig{figure=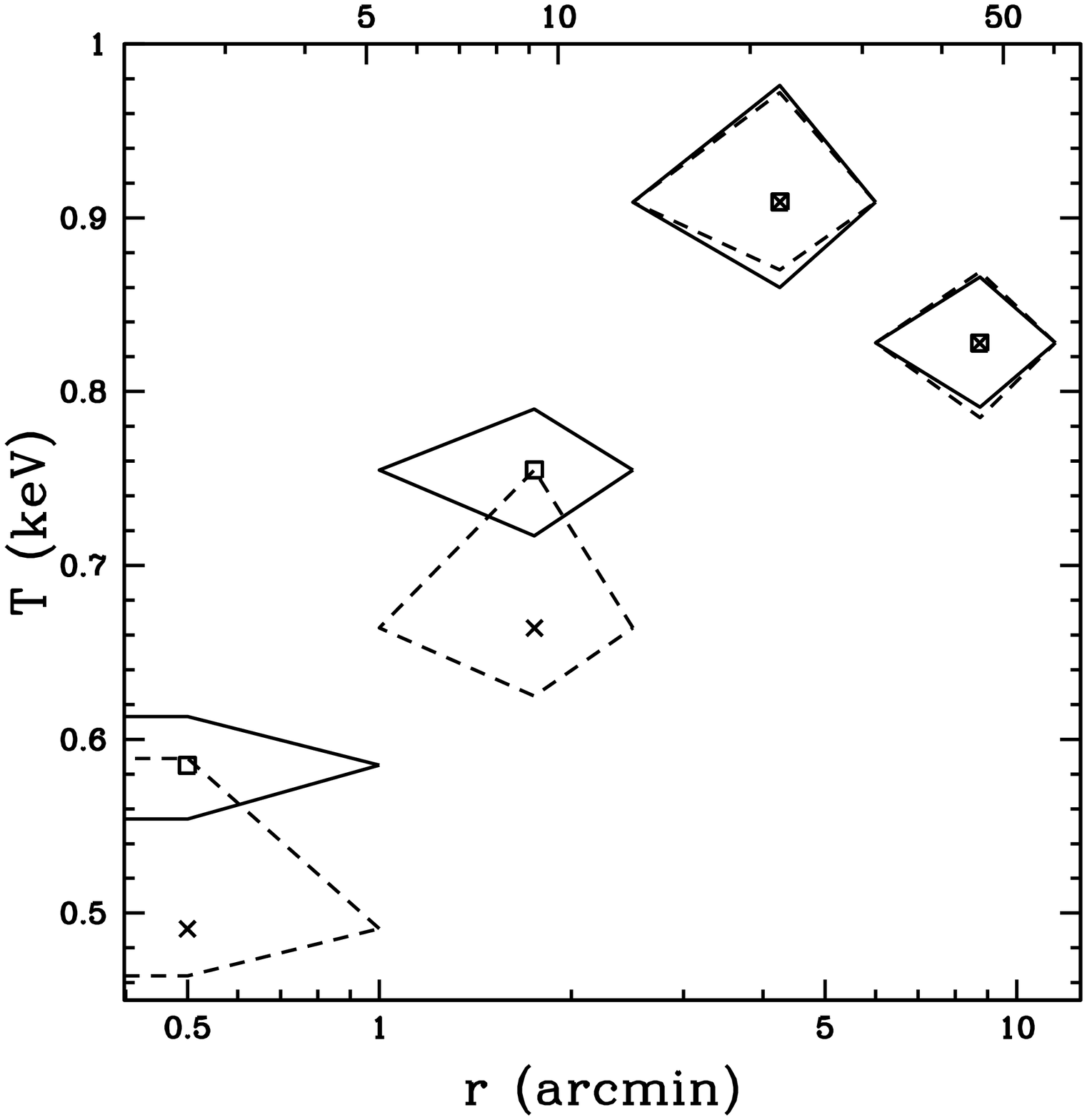,height=0.22\textheight}}
}
\parbox{0.32\textwidth}{
\centerline{\psfig{figure=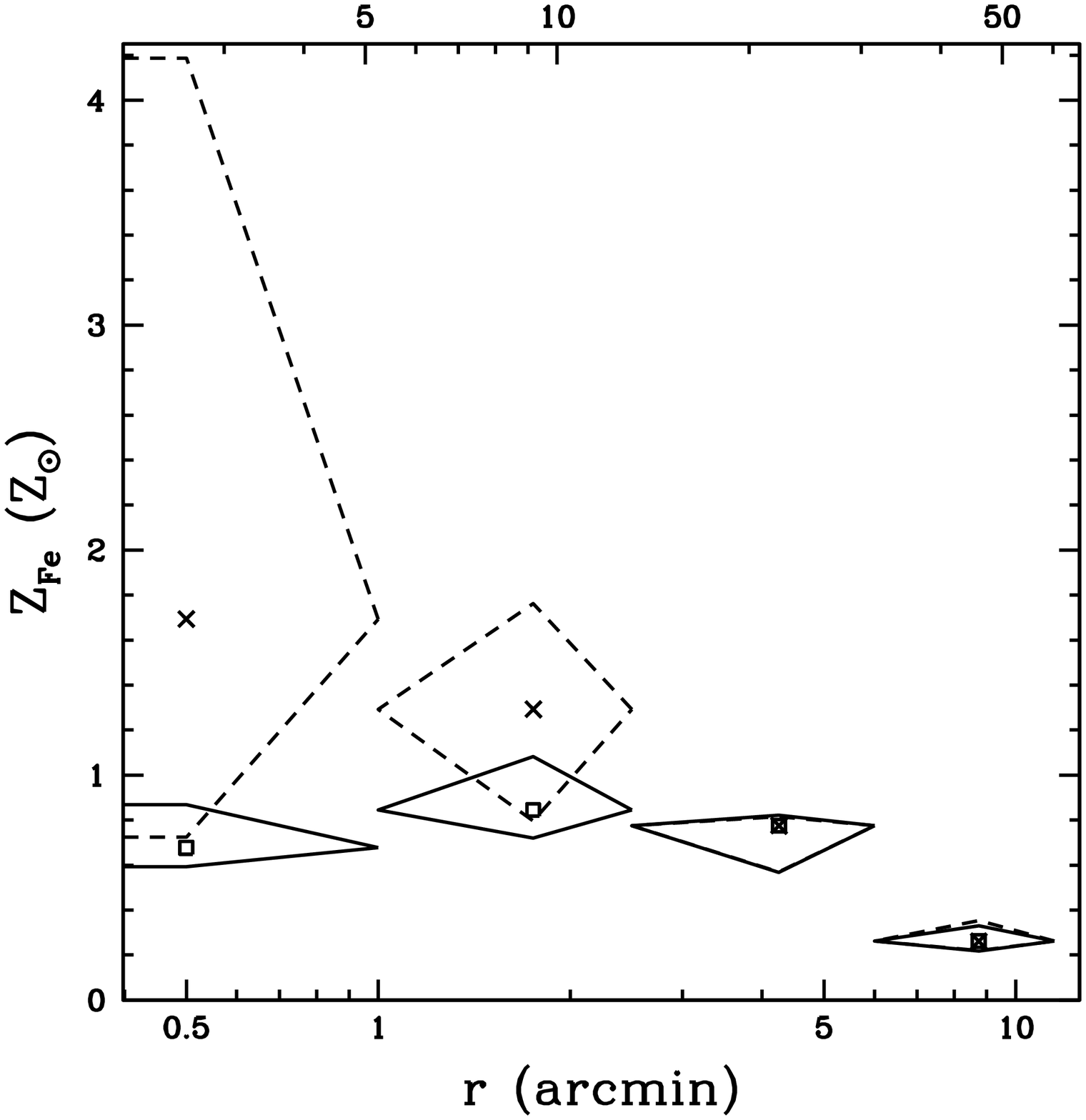,height=0.22\textheight}}
}
\parbox{0.32\textwidth}{
\centerline{\psfig{figure=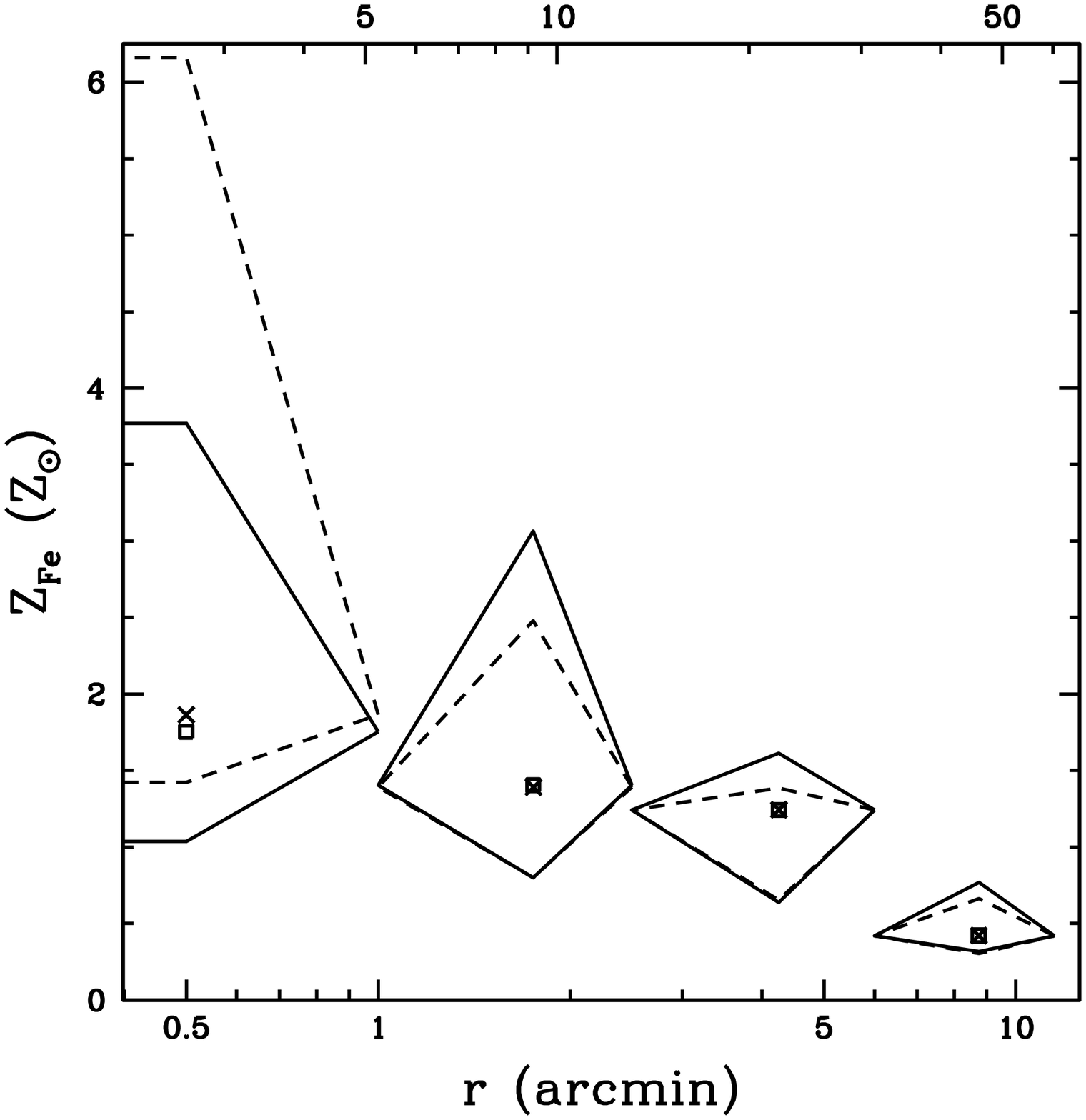,height=0.22\textheight}}}
\vskip 0.25cm
\centerline{\large\bf HCG 62} \vskip 0.1cm
\parbox{0.32\textwidth}{
\centerline{\psfig{figure=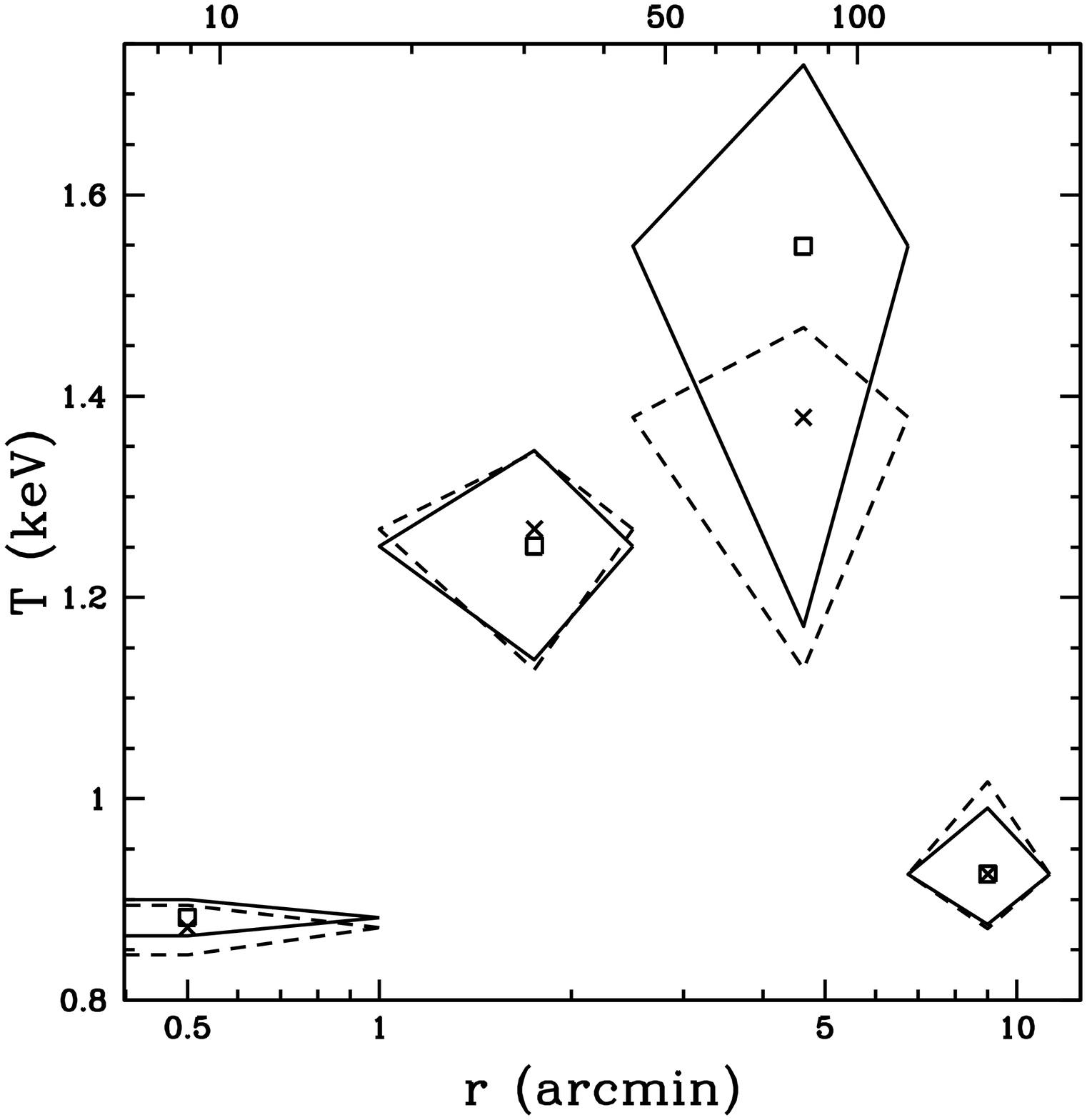,height=0.22\textheight}}
}
\parbox{0.32\textwidth}{
\centerline{\psfig{figure=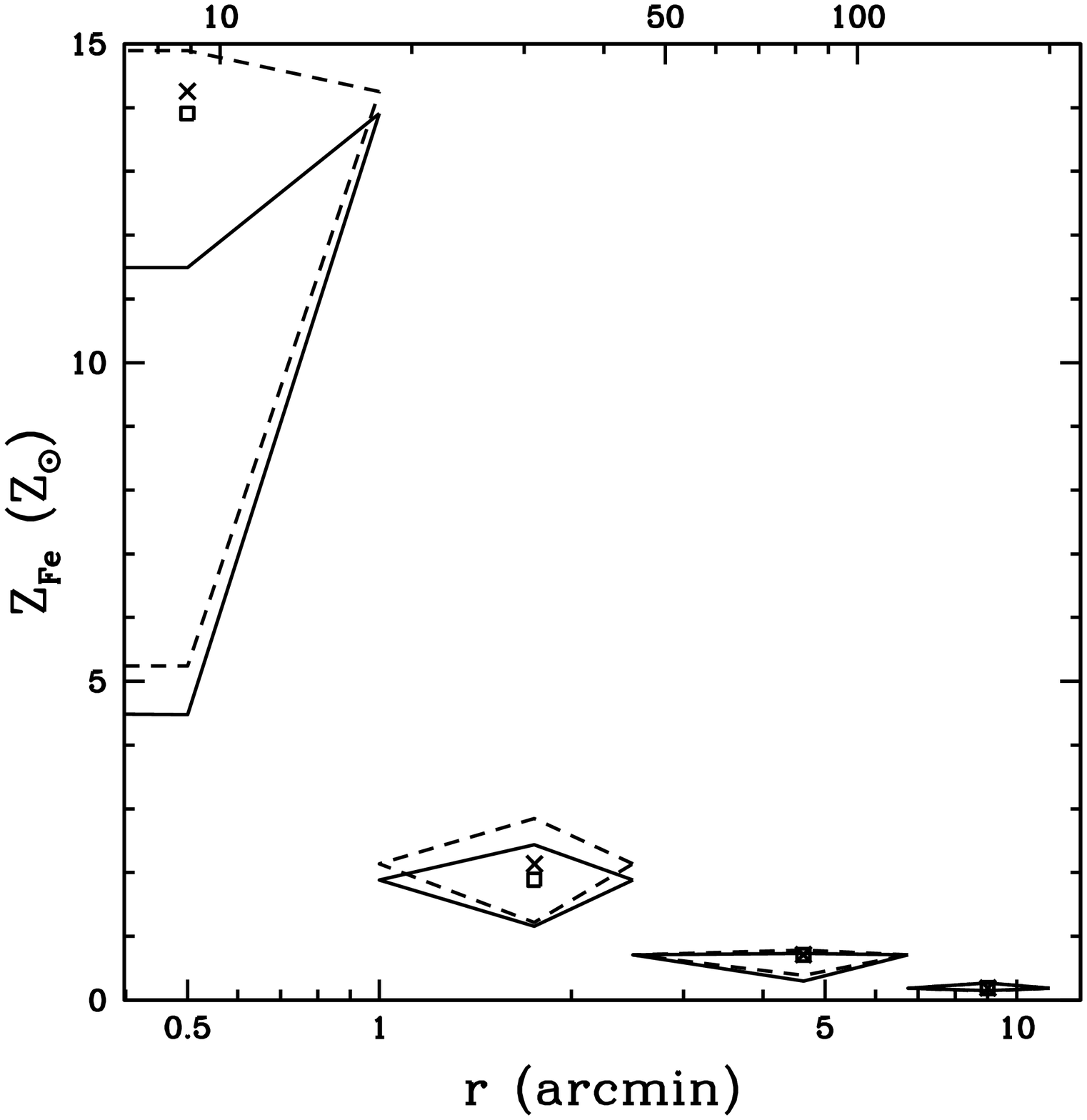,height=0.22\textheight}}
}
\parbox{0.32\textwidth}{
\centerline{\psfig{figure=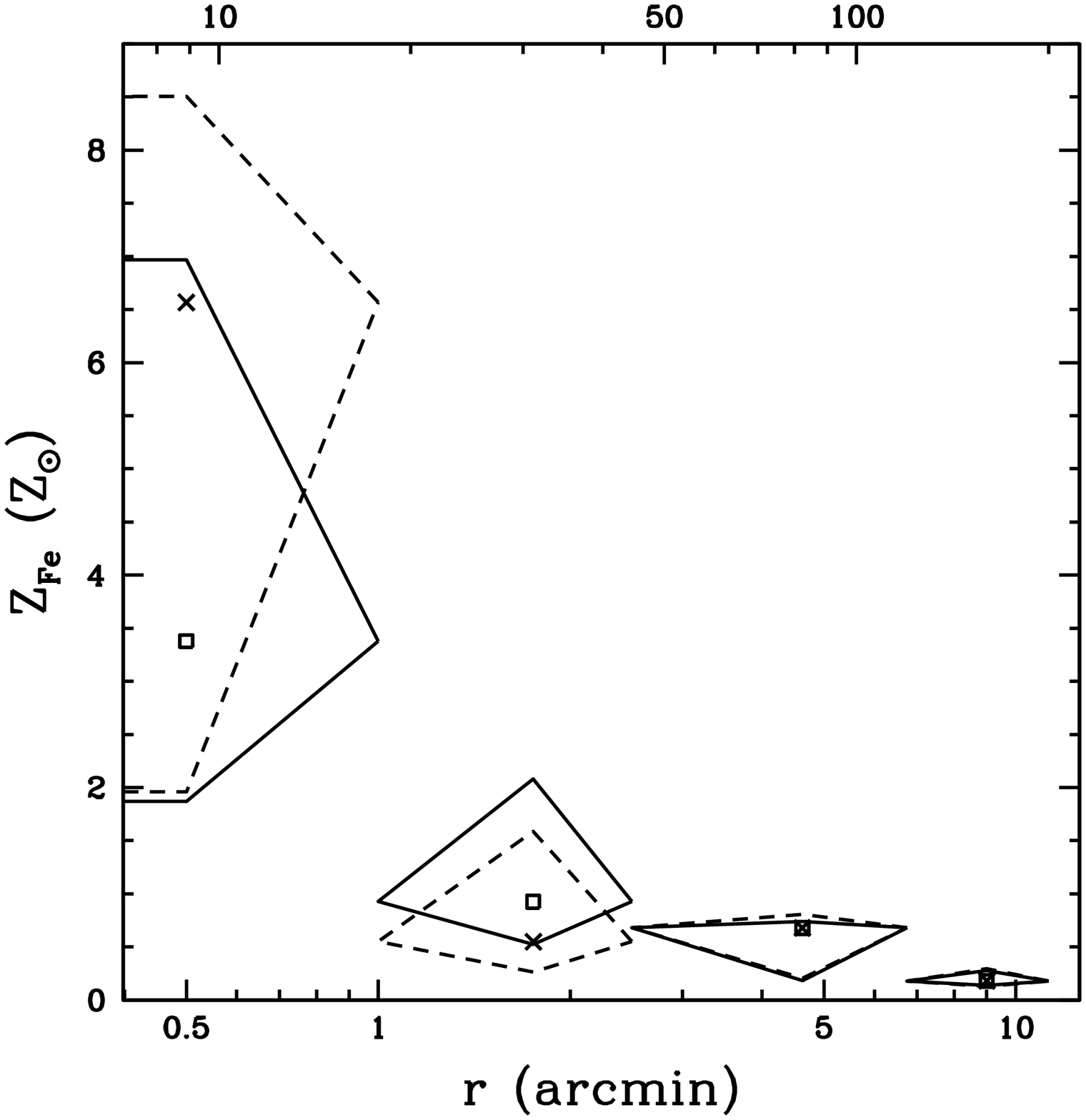,height=0.22\textheight}}}

\caption{\label{fig.4ann} As Figure \ref{fig.7ann} but for systems
with 4 annuli.}
\end{figure*}

\begin{figure*}[t]
\centerline{\large\bf NGC 4649} \vskip 0.1cm
\parbox{0.32\textwidth}{
\centerline{\psfig{figure=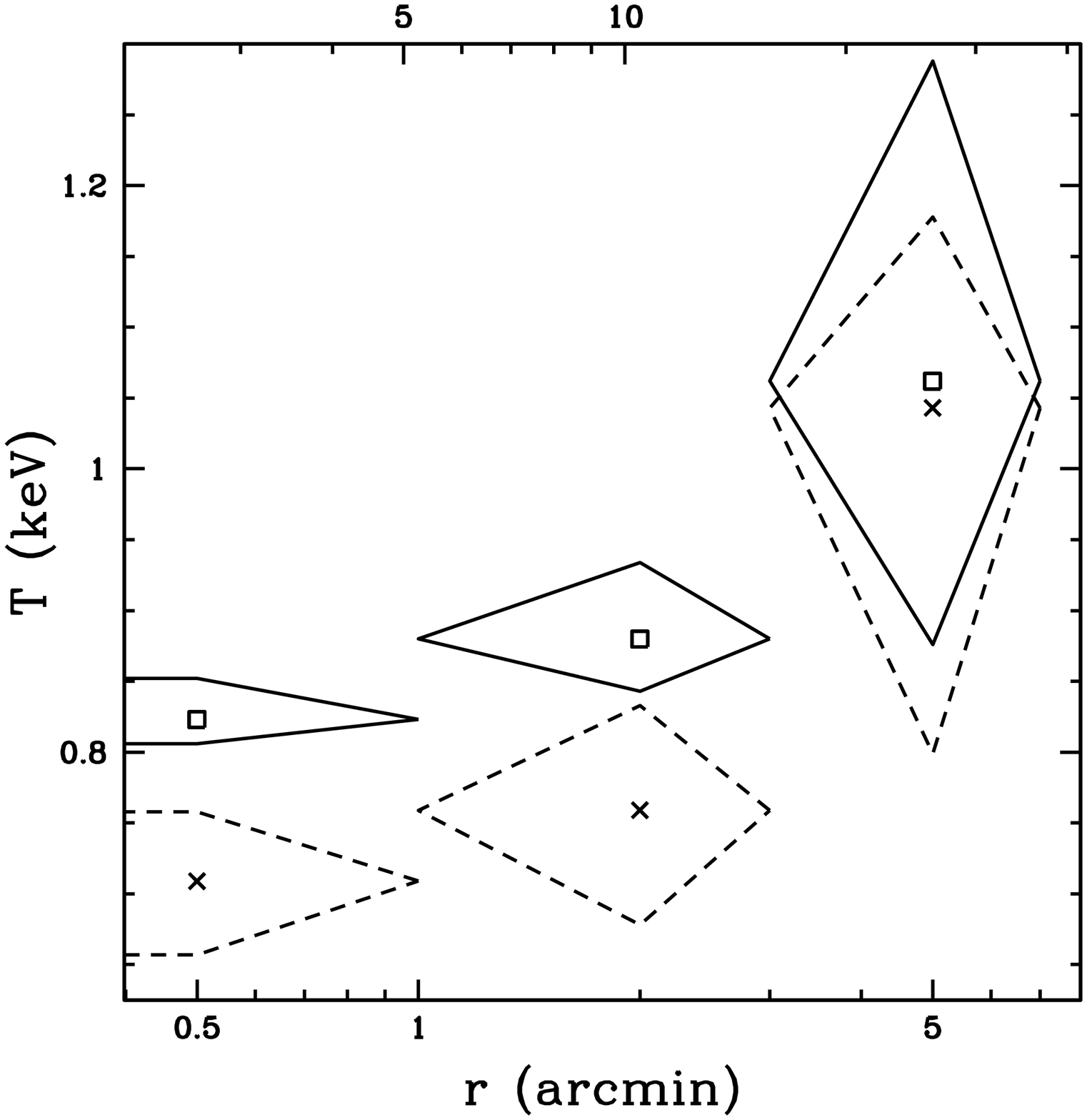,height=0.22\textheight}}
}
\parbox{0.32\textwidth}{
\centerline{\psfig{figure=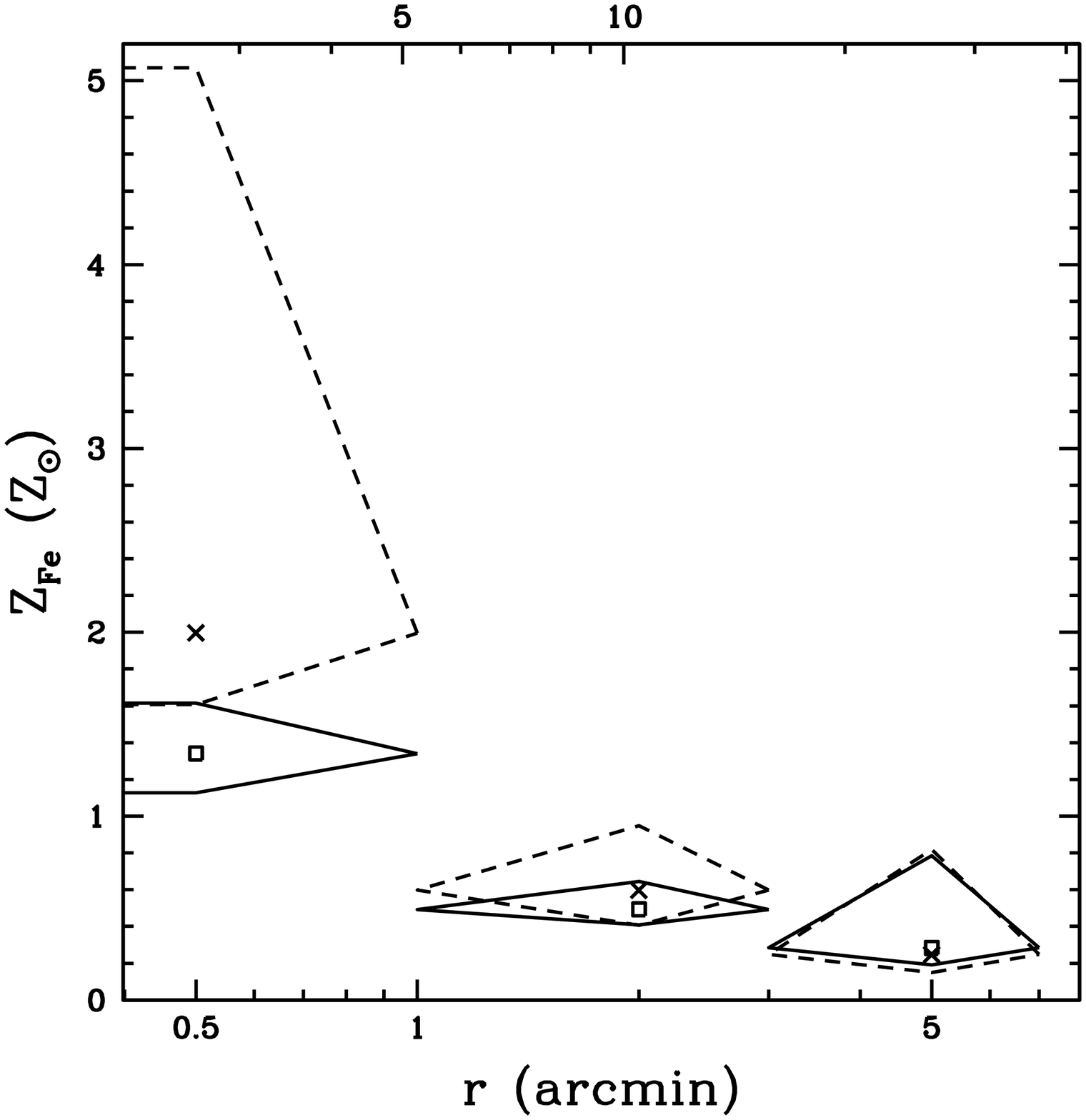,height=0.22\textheight}}
}
\parbox{0.32\textwidth}{
\centerline{\psfig{figure=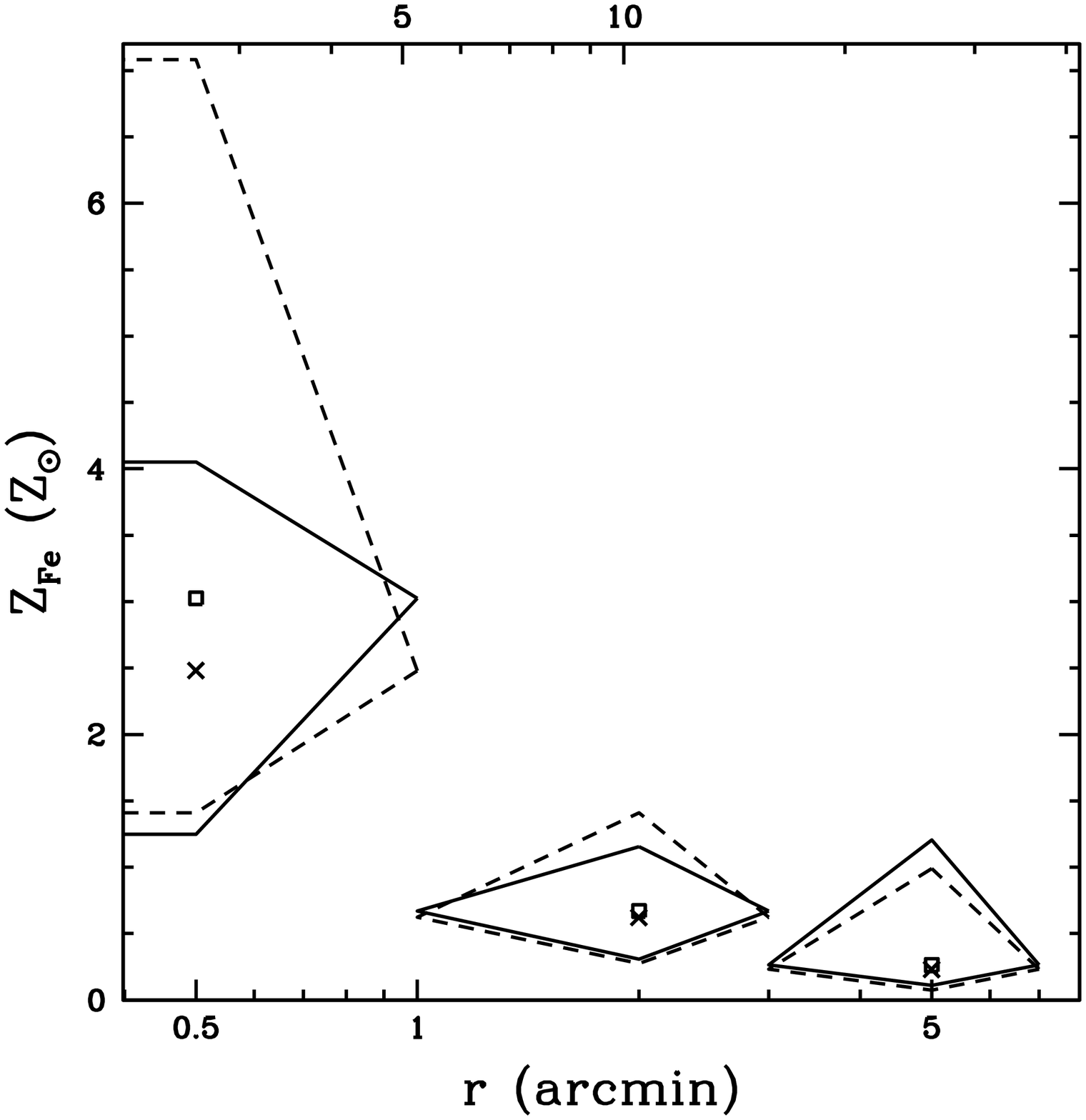,height=0.22\textheight}}}
\caption{\label{fig.3ann} As Figure \ref{fig.7ann} but for the one system
with 3 annuli.}
\end{figure*}

We plot in Figures \ref{fig.7ann}-\ref{fig.3ann} the temperature and
Fe abundance profiles obtained from the deprojection analysis
according to the number of annuli for which useful constraints on the
parameters were obtained. This categorizes the systems essentially
according to the S/N of the data. Fe abundance profiles are shown for
the cases where the column density of the standard absorber was (1)
fixed at the Galactic value and (2) treated as a free parameter. (The
temperature profiles are very similar in each case, and thus for
clarity of presentation we show only the case for fixed Galactic
column density.)  For both the temperature and Fe abundance profiles
we also show the results obtained with and without an extra oxygen
edge at 0.532 keV (rest frame). The profiles of column density and
edge optical depth are given in PAPER1 for NGC 1399 and 5044 and in
PAPER3 for all the systems.

The deprojected temperature profiles are all very similar to the 2D
profiles obtained in many previous studies with the \rosat PSPC
(Forman et al 1993; Ponman \& Bertram 1993; David et al 1994;
Trinchieri et al 1994; Rangarajan et al 1995; Kim \& Fabbiano 1995;
Irwin \& Sarazin 1996; Jones et al 1997; Trinchieri et al 1997;
Mulchaey \& Zabludoff 1998; Buote 1999). As expected, our 3D profiles
display a slightly steeper rise from the central minimum to the
maximum at $r\sim 50$-100 Mpc in very good agreement with the 3D
profiles obtained by Finoguenov and co-workers for several systems
using both \rosat and \asca data (Finoguenov et al 1999; Finoguenov \&
Ponman 1999; Finoguenov \& Jones 2000). The temperature profiles also
agree very well with two-temperature models of \asca data (Buote 1999,
2000a).

The models with an oxygen edge tend to have lower temperatures at
small radii. This can occur because lower temperature gas produces
stronger oxygen lines which are absorbed by the oxygen edge. Although
in most cases both models give fits of similar quality, the edge model
provides a significantly better fit for some of the systems (see
PAPER1 and PAPER3).

For the four systems with the highest S/N data in our sample (NGC 507,
1399, 4472, and 5044) the Fe abundance profiles obtained from models
(without an oxygen edge) where the column density is treated as a free
parameter are very much larger than for all models with fixed Galactic
column density. Even after accounting for the differences between
meteoritic and photospheric solar abundances, the Fe abundances
implied by these models greatly exceed and are very inconsistent with
the Fe abundances obtained from all \asca studies. They are, however,
consistent with previous \rosat studies where the column density was
treated as a free parameter (e.g., Forman et al 1993; David et al
1994).

For these systems the models with variable column density give
Fe abundances that increase rapidly with decreasing radius, but the
column densities (except for NGC 507) also {\it decrease} well below
the Galactic values as $r\rightarrow 0$ (PAPER1 and PAPER3).  Since
the column densities at large radii are consistent with the Galactic
values this decrease at small radii cannot be due to errors in the
background subtraction. Rather, the sub-Galactic columns and huge
Fe abundances that are inconsistent with \asca signals some inadequacy
in the emission model. As it turns out (see Figures
\ref{fig.7ann}-\ref{fig.3ann}) except for NGC 5044 the variable-column
density models with an oxygen edge give Fe abundances (and columns)
that are very consistent with the fixed-column models. This gives
additional support for the edge model discussed in PAPER1 and PAPER3
(though see \S 5.3 of PAPER3).

In most of the other systems the variable-column-density models also
tend to predict sub-Galactic columns at small radii though at much
lower significance. These sub-Galactic columns are also accompanied by
systematically larger Fe abundances than the fixed-Galactic case,
though again the edge models usually give consistent abundances for
the free- and fixed-column density cases. Hence, the Fe abundance
profiles which typically agree best with previous \asca analyses are
those obtained from models with fixed Galactic column density (with or
without an edge), though variable column models with an oxygen edge
yield comparable results in most cases. Since, however, the evidence
for intrinsic oxygen absorption is strong for half of these systems
(PAPER1 and PAPER3) we believe the models with an intrinsic oxygen
edge and a standard absorber with $\nh = \nhgal$ to be the most
physical of the models investigated (i.e. crosses and dotted diamonds
in middle column of Figures \ref{fig.7ann} - \ref{fig.3ann}).

From inspection of Figures \ref{fig.7ann}-\ref{fig.3ann} it is readily
apparent that most of the systems have Fe abundance profiles that
decrease as a function of radius. Typically within $r\sim 10$ kpc the
Fe abundance ranges from $\fe\approx 1\solar - (\rm several)\solar$ and
decreases to $\fe\sim 0.5$\solar for $r\sim 50-100$ kpc. We now discuss
the individual systems in groups defined by the total number of annuli
with interesting constraints. When comparing our Fe abundances to
previous studies we implicitly account for projection effects and the
different plasma codes and solar abundances used (\S \ref{solar}).

\subsubsection{Systems with 7 Annuli}

In Figure \ref{fig.7ann} we display the results for the three systems
where the spectral parameters are well determined in seven
annuli. These observations thus generally correspond to the highest
S/N data in our sample. In each case the models with an oxygen edge
exhibit significant negative Fe abundance gradient, though in NGC 1399
the key evidence occurs in the central radial bin.

\medskip
\noindent {\bf NGC 507:} The evidence for a Fe abundance gradient is
strong as is the large value in the central bin: 95\% confidence lower
limits are 1.3\solar\thinspace and 1.1\solar\thinspace respectively
for models with and without an oxygen edge (column density fixed at
Galactic). The average Fe abundance within $r=4\arcmin$ is $\approx
1\solar$ in excellent agreement with the value obtained by \citet{bf}
from a two-temperature model of the \asca data. Our Fe abundance
profile also agrees reasonably well with the 2D profile obtained in
the previous \rosat study by \citet{kf}.

\noindent {\bf NGC 1399:} The Fe abundance profile is constant for
$r\ga 3\arcmin$, rises slightly for $r\approx 1\arcmin$-$2\arcmin$,
and in the center has different values depending on the model. As
shown in PAPER1 and PAPER3 the edge model is strongly preferred, and
thus the Fe abundance is consistent with being largest at the center;
note that even though the error estimate for the fixed-column-density
case appears to underestimate the Fe abundance, we still estimate a
95\% lower limit of 1.1\solar which is significantly greater than at
large radii. The average Fe abundance within $r=4\arcmin$ is $\approx
1.4\solar$ which agrees with the lower limit obtained by \citet{b99}
from a two-temperature model of the \asca data and is in better
agreement with a cooling flow model. Our Fe abundance profile for the
case of variable column density and no edge agrees very well with the
2D profile of \citet{j97}.

\noindent {\bf NGC 5044:} Again there is clear evidence that the
Fe abundance profile is not constant with radius. As noted earlier the
models with variable column density predict large Fe abundances that
are inconsistent with \asca studies. (But they are consistent with the
previous \rosat study of David et al 1994.) For example, \citet{b99}
obtains $\fe\approx 0.9\solar$ within $r\approx 5\arcmin$ for a
two-temperature model of the \asca data. Since in PAPER1 and PAPER3 we
also find that an oxygen edge is clearly required in the central
regions, let us focus on the Fe abundance profile obtained from a model
with an edge and with fixed Galactic column density. This model yields
a Fe abundance of $\fe\sim 0.7\solar$ for $r\ga 3\arcmin$ and $\fe\sim
1.5\solar$ for smaller $r$. 

\subsubsection{Systems with 5-6 Annuli}

The results for systems having 5 or 6 annuli are plotted in Figure
\ref{fig.6ann}. Evidence for a negative Fe abundance gradient is clear
for NGC 4472, marginal for NGC 5846, and non-existent for NGC 2563.

\medskip
\noindent {\bf NGC 2563:} Independent of the model it is clear that
the Fe abundance is constant with radius with a value of $\sim
2.0\solar$ when the column density is treated as a free parameter as
opposed to $\sim 1.5\solar$ for fixed Galactic column density. These
results are consistent with the Fe abundance determined within
$r\approx 3\arcmin$ by \citet{b00a} from a two-temperature model of
the \asca data.

\noindent {\bf NGC 4472:} Analogously to NGC 5044 the Fe abundances
obtained for the model with variable column density and no edge are
very large and very inconsistent with all previous \asca
measurements. Also similar to NGC 5044 is that the edge model is
strongly preferred (see PAPER3). The edge models give $\fe\la 1\solar$
for $r>3\arcmin$ and $\fe\sim 2\solar-3\solar$ at smaller radii. These
values are in excellent agreement with those obtained from previous 2D
studies with \rosat (Forman et al 1993; Irwin \& Sarazin 1996). These
Fe abundances also agree very well with the value obtained by
\citet{b99} from a two-temperature model of the \asca data.

\noindent {\bf NGC 5846:} The models with variable column density show
clear evidence for a negative Fe abundance gradient, but since the
column densities obtained within the inner two bins are sub-Galactic
we do not consider these models to be physical (\S 5.3 of PAPER3). The
models with fixed Galactic column density with and without an edge
give very similar profiles except that in the central bin the upper
limit on the Fe abundance for the edge model is not very well
determined. These fixed-column models suggest a negative gradient in
Fe abundance, but the errors are sufficiently large so that a constant
profile is not clearly excluded. These Fe abundances also agree very
well with the values obtained by \citet{b00a} from multitemperature
models of the \asca data.

\subsubsection{Systems with 4 Annuli}

The galaxies and groups for which we obtained useful constraints in
four radial bins are shown in Figure \ref{fig.4ann}. Although the data
sets for these systems have lower S/N than the previous examples, the
evidence for negative Fe abundance gradients is as strong or stronger
than the others.

\medskip
\noindent {\bf NGC 533:} All models give $\fe\sim 0.5\solar$ for $r\ga
3\arcmin$ and $\fe\ga 1\solar$ within $1\arcmin$. These
Fe abundances are consistent with the results for multitemperature
models obtained with \asca within $r=3\arcmin$ by \citet{b00a}.

\noindent {\bf NGC 4636:} In the outer bin all models give a very
sub-solar value for the Fe abundance, $\fe\sim 0.4\solar$. The
Fe abundance rises with decreasing radius such that $\fe\ga 1\solar$ in
the central bin. These results are consistent with those obtained by
\citet{b99} for multitemperature models of \asca data analyzed within
$r=5\arcmin$.

\noindent {\bf HCG 62:} Similar to NGC 4636, all models give a very
sub-solar value for the Fe abundance, $\fe\sim 0.2\solar$, in the outer
bin. The Fe abundance rises very sharply with increasing radius such
that $\fe\ga 2\solar$ in the central bin. This system appears to possess
the most significant Fe abundance gradient in our sample: the 95\%
lower limits on the values in the central bin are 4.6\solar\thinspace
and 3.2\solar\thinspace respectively for models with and without an
edge (both with column density fixed to Galactic). The Fe abundance of
$\sim 1.4\solar$ obtained by \citet{b00a} using a two-temperature
model of the \asca data accumulated within $r=3\arcmin$ is consistent
with our \rosat results provided the Fe abundance in the central bin is
near the estimated $1\sigma$ lower limits.

\subsubsection{Systems with 3 Annuli}

Finally, in Figure \ref{fig.3ann} we show the results for the galaxy
with the smallest number of annuli, NGC 4649. Similar to the 4-annuli
objects, NGC 4649 is one of the most significant examples for a
decrease in Fe abundance with increasing radius.

\medskip
\noindent {\bf NGC 4649:} For this system the edge model is clearly
preferred (see PAPER3), though the Fe abundances are fairly similar
for all of the models. In the outer bin $\fe\sim 0.25\solar$ for all
models and rises to $\fe\ga 1.5$ within the central bin for the edge
models. \citet{bf} obtained a Fe abundance of $\sim 1.3\solar$ from a
two-temperature model of the \asca data within $r=3\arcmin$ in good
agreement with our \rosat results.

\subsection{Caveats}
\label{caveats}

\noindent {\it (i) Spherical Symmetry:} The X-ray isophotes of most of
the systems are approximately circular over the regions
examined. Notable exceptions are NGC 1399 and 4472 which show
significant deviations from circular symmetry at large radius. For
$R\ga 5\arcmin$ there is a significant N-S asymmetry in the X-ray
surface brightness of NGC 1399 which \citet{j97} hypothesize is due to
incomplete relaxation of the gas in the surrounding group. In
contrast, outside of $R\ga 3\arcmin$ the X-ray isophotes of NGC 4472
flatten which \citet{is} attribute to ram pressure as the galaxy moves
with respect to the Virgo cluster. Our results for these systems at
large radii average over these asymmetries. It is worth remarking that
although these systems have larger than solar Fe abundances at their
centers, we find that at the largest radii $\fe\sim 0.75\solar$ for NGC
1399 and $\fe\sim 0.3\solar$ for NGC 4472; i.e. it is plausible that the
small value for NGC 4472 is significantly influenced by the ambient
gas of the Virgo cluster.

\noindent {\it (ii) Single-Phase Gas:} We have focused on single-phase
analysis of the hot gas because in no case did we find that multiphase
models of the hot gas improved the fits significantly. In the central
bins of most of the systems we expect gas to be emitting over a range
of temperatures because of the observed temperature
gradients. However, deprojection has removed the high-temperature gas
components from the centers leaving a much smaller temperature range
(typically a few tenths of a keV) which cannot be distinguished by the
limited energy resolution of the PSPC. (Note that even if gas is
dropping out of a cooling flow, emission weighted temperatures are
within 50\% of the ambient value -- e.g., Buote, Canizares, \& Fabian
1999). These small temperature ranges also indicate that the ``Fe
Bias'' should be unimportant (see appendix of Buote 2000a), and thus
multitemperature models do not give qualitatively different Fe
abundances in the central bins -- see also \S \ref{bias}.

\noindent {\it (iii) Background Issues:} There are two backgrounds
which must be considered. First, the edge effect described in \S
\ref{edge} requires one to assume a model for the emission outside of
the bounding annulus. We never found any noticeable effect on the
derived temperatures and abundances for reasonable choices of
$\beta$. Only the electron density is affected and only for the
outermost annuli. Second, we examined the effects of taking background
estimates from different parts of the detector and from different
fields nearby to a given object. We found that in most cases when the
background level was over-estimated the Fe abundances tended to be
larger than when the background was under-estimated. The size of the
effect depends on many factors (e.g., S/N) though again we found that
for reasonable background choices only the outermost annuli are
affected as would be expected. (Of course such effects will propagate
to smaller radii if the regularization criteria are not applied with
care -- see below.)

\section{Discussion and Conclusions}
\label{disc}

Deprojection analysis of the \rosat PSPC data of 10 cooling flow
galaxies and groups reveals clear evidence for Fe abundances that
decrease with radius in all but one system (NGC 2563). Typically
$\fe\sim 0.5\solar$ at the largest radii examined ($r\sim 50$-100 kpc)
which increases to $\fe\approx 1\solar - (\rm several)\solar$ within the
central radial bin ($r\la 10$ kpc). In most cases the estimated
uncertainties on $\fe$ are large within the central bin and allow for
$\fe\gg 1\solar$ (but never $\fe\ll 1\solar$). Throughout this paper we
have used the ``meteoritic'' solar abundances (Fe/H is $3.24\times
10^{-5}$ -- see Ishimaru \& Arimoto 1997) which leads to Fe abundances
that are a factor of 1.44 larger than previous studies that used the
old ``photospheric'' solar value for Fe (see \S \ref{solar}).

These 3D Fe abundance profiles are generally consistent with the
original 2D \rosat studies (Forman et al. 1993; David et al 1994; Kim
\& Fabbiano 1995; Irwin \& Sarazin 1996; Jones et al 1997) after
accounting for projection effects and for the different plasma codes
and solar abundances used (though see below in \S \ref{bias}). For all
10 systems the Fe abundances (and temperatures) that we have obtained
from the deprojected \rosat data agree with those we have obtained
previously from analysis of the \asca data accumulated within $r\sim
3\arcmin$-$5\arcmin$ (Buote \& Fabian 1998; Buote 1999, 2000a);
qualitatively similar results for some of these systems also have been
recently obtained with \asca data by Allen, Di Matteo, \& Fabian
(2000).

Therefore, within $r\sim 50$ kpc of these bright galaxies and groups
the \rosat and \asca data clearly demonstrate that the gas is
non-isothermal with approximately solar Fe abundances. We now examine
whether it is the assumption of isothermality or the presence of Fe
abundance gradients which accounts for the very sub-solar Fe
abundances inferred by most previous \asca (and some \rosat) studies
(see Buote 2000a for a detailed review).

\subsection{Fe Bias vs Fe Gradients}
\label{bias}

\begin{figure*}[t]
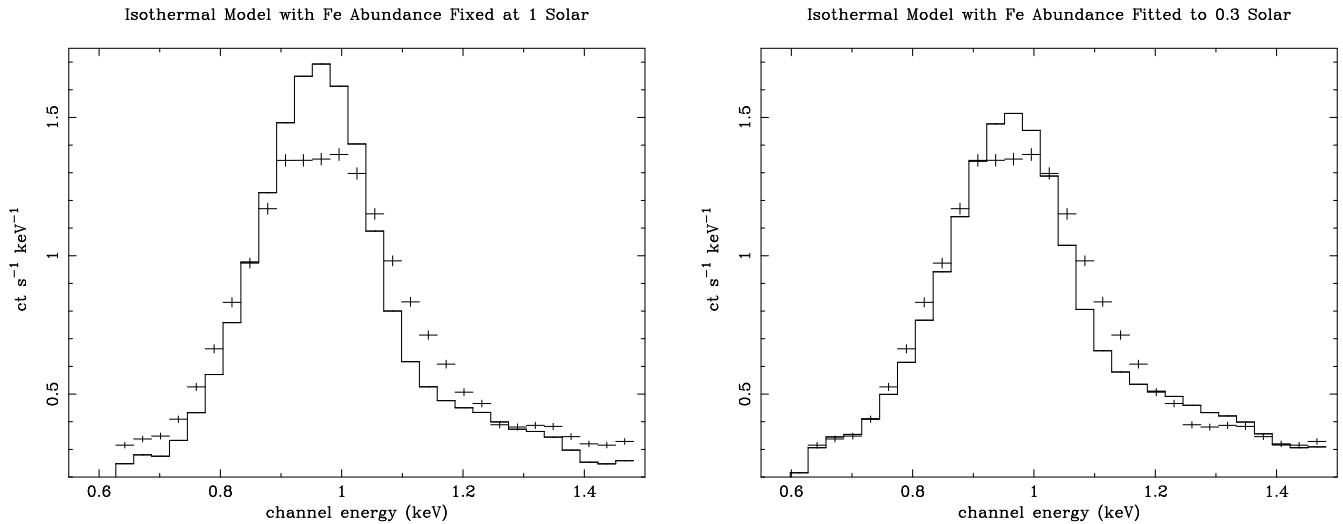

\parbox{0.49\textwidth}{
\centerline{\psfig{figure=fe_bias_fix.ps,angle=-90,height=0.28\textheight}}
}
\parbox{0.49\textwidth}{
\centerline{\psfig{figure=fe_bias_free.ps,angle=-90,height=0.28\textheight}}
}
\caption{\label{fig.bias} (Left) Simulated \asca SIS spectrum of a
two-temperature plasma ($T_1=0.75$ keV, $T_2=1.5$ keV with equal
emission measures for each temperature component) with $\fe=1\solar$
(crosses) and the best-fitting single-temperature model (solid line)
also with \fe\, fixed at $1\solar$. (Right) Same but the Fe abundance
of the single-temperature model is allowed to vary resulting in a
best-fitting value of $\fe=0.3\solar$.}
\end{figure*}

Previously we have discussed how fitting a single-temperature plasma
model to the soft X-ray spectrum of a bright elliptical galaxy or
galaxy group that actually consists of multiple temperature components
can lead to a serious underestimate of the Fe abundance (Buote \&
Canizares 1994; Buote \& Fabian 1998; Buote 1999, 2000a).  In our most
comprehensive treatment to date \citep{b00a} we provided a physical
explanation of this ``Fe Bias'' from examination of \asca data which
we now summarize in Figure \ref{fig.bias}.

The accumulated \asca spectra within $r\sim 3\arcmin$-$5\arcmin$ of
the brightest elliptical galaxies and groups (including those in our
present investigation) are well represented by two-temperature (2T)
models. These 2T models are also consistent with the radial
temperature gradients inferred from the \rosat data within the same
spatial regions (Buote 1999, 2000a). Moreover, we have verified that
such 2T models continue to provide a good representation of the
accumulated spectra generated by the \rosat temperature profiles out
to the largest radii ($\sim 15\arcmin$) investigated in our present
paper whether or not the temperature rises and then falls (e.g., NGC
2563 -- Figure \ref{fig.6ann}) or continues to rise out to the largest
radii (e.g., NGC 4472 -- Figure \ref{fig.6ann}). This agreement is
expected because a 2T model can mimic very accurately the soft X-ray
spectra of cooling flows \citep{bcf}.

In Figure \ref{fig.bias} we show a simulated \asca SIS spectrum of a
2T model with $\fe=1\solar$ with temperatures and relative emission
measures typical of the bright galaxies and groups.  For this
discussion we consider only the energies 0.55-1.5 keV to emphasize the
Fe L lines and the lower bandpass limit of the SIS.  Also shown in
Figure \ref{fig.bias} is the result of fitting a single-temperature
(1T) model with \fe\, fixed at $1\solar$ to the simulated 2T spectral
data. It is readily apparent that the 1T model is a poor fit to the 2T
spectrum as it is too peaked near 1 keV while being deficient in
emission at other energies. This behavior simply reflects the ability
of the 2T model to excite Fe L lines over a wider range of energies
than the 1T model. In order to force the 1T model to better fit the 2T
spectrum one has to reduce the size of the peak at 1 keV which means
reducing the Fe abundance. Allowing \fe\, to be a free parameter
results in a better (but still not good) fit and with a best-fitting
value of $\fe=0.3\solar$ (see Figure \ref{bias}). It is this effect
that we have termed the ``Fe Bias''.

This Fe Bias is dependent on the energy resolution of the detector
and is more pronounced at higher resolution which is why we have
focused on the \asca SIS data. But if the models in Figure \ref{bias}
are folded with the relatively low resolution of the \rosat PSPC we
obtain a best-fitting value of $\fe=0.4\solar$ for the 1T model, very
similar to that obtained above at higher resolution with the
SIS. However, it must be emphasized that this underestimate with the
PSPC is achieved only if the lower limit on the bandpass is restricted
to energies $\ga 0.6$ keV as appropriate for the SIS. If instead the
lower energy limit is reduced to 0.2 keV as is appropriate for the
PSPC then we obtain $\fe\approx 0.6-0.7\solar$ for the 1T model.

This smaller underestimate occurs because while reducing the Fe
abundance diminishes the line emission of the 1T model to better fit
the spectral maximum near 1 keV, the continuum of the 1T model must
increase to compensate for the decrease in the Fe L emission at
energies near $\sim 0.8, 1.2$ keV. It also compensates for the
corresponding decrease in the strong O K$\alpha$ emission near 0.6 keV
since O (like the other elements) varies with Fe in its solar ratio
(see \S \ref{solar}). 

This raising of the continuum by the 1T model generally goes unnoticed
when analyzing \asca data for the following reason. The \asca SIS and
GIS bandpasses do not extend below the energy regions shown in Figure
\ref{fig.bias}, but they do include energies up to $\sim 10$
keV. However, the emission at higher energies is usually accounted for
by introducing an additional high-temperature bremsstrahlung component
(BREM) to represent emission from discrete sources (e.g., Matsumoto et
al. 1997), and thus any increase or decrease in the emission of the 1T
model at energies much larger than 1 keV can be compensated by
changing the normalization of the BREM component. The required amount
of compensation is generally unimportant because the emission from the
1T model decreases as $\exp(-E/T)$, and thus a plasma with $T\approx
1$ keV contributes little to the emission of a 1T+BREM model at higher
energies.

However, spectral analysis with PSPC data includes lower energies
(i.e. $\sim 0.2-0.5$ keV) that are dominated by continuum emission.
The data at these energies restrict the attempt by the data near 1 keV
to simultaneously reduce the Fe abundance and increase the continuum
in order to force a better fit of a 1T model to the 2T spectrum.
Hence, the Fe Bias affecting analysis of PSPC data is much less
pronounced than for \asca data because of the lower energy resolution
of the PSPC, and especially because the PSPC data include energies
below $\sim 0.5$ keV.

Our deprojection analysis of the PSPC data removes the temperature
components projected from larger radii from the spectrum of a given
annulus. Consequently, if the gas is single-phase (i.e., a single
density and temperature at each radius) the range of temperatures
within a given annular spectrum must be small after deprojection
(i.e., essentially the temperature difference between the annuli that
surround the annulus in question). Since each deprojected spectrum is
nearly isothermal, and since the Fe Bias is already greatly reduced
when analyzing PSPC data, we expect that the Fe Bias does not
significantly affect our measurements of the Fe abundances from the
PSPC data in this paper. It is therefore not surprising that the Fe
abundances we have obtained from our deprojection analysis of the PSPC
data agree only with those obtained from the \asca studies that have
removed the Fe bias by fitting multitemperature models of the \asca
spectra accumulated within substantial spatial regions ($r\sim
3\arcmin - 5\arcmin$; Buote \& Fabian 1998; Buote 1999, 2000a; Allen
et al 2000).

If a warm gas phase ($T=10^5-10^6$ K) also contributes to the X-ray
emission of the galaxies and groups as indicated by the intrinsic
absorption reported in PAPER1 and PAPER3, then the single-phase
assumption needs to be reconsidered. However, the low temperature of
the warm gas implies a negligible contribution to the Fe L emission
and thus will not contribute to an Fe Bias. Another possibility is
that the hot gas emits over a continuous range of temperatures as
might be expected from an inhomogeneous cooling flow that cools at
constant pressure (e.g., Johnstone et al 1992). As mentioned in \S
\ref{caveats} we do not find any significant change in the inferred Fe
abundances if we use such a cooling flow model.

If the Fe Bias is not very important for analysis of \rosat PSPC data,
then why do some \rosat studies find very low Fe abundances? As stated
above the Fe abundances we have determined from the deprojection
analysis of the PSPC data of the 10 galaxies and groups in our sample
agree very well with those obtained from previous spatial-spectral
analyses of these systems with the PSPC after accounting for (1) the
factor of 1.44 arising from our use of the meteoritic solar
abundances, (2) different plasma codes used, and (3) small differences
associated with our use of a deprojection analysis (Forman et
al. 1993; David et al 1994; Kim \& Fabbiano 1995; Irwin \& Sarazin
1996; Jones et al 1997).

Our results also agree with those of \citet{hp} who fitted 1T models
within large apertures (typically $20\arcmin$ radius) for several
groups including NGC 533, 2563, 4636, and 5846.  To make a more
realistic comparison we summed up the emission models we obtained for
all annuli for these systems, simulated a PSPC observation using the
same exposure time, and then fitted the simulated spectrum with a 1T
model like Helsdon \& Ponman. The Fe abundances we obtained from this
exercise are fully consistent within the $2\sigma$ errors for each
system. We emphasize that Helsdon \& Ponman find $\fe\ga 0.6\solar$
(scaled to meteoritic solar) for these systems even when using very
large apertures.

In contrast \citet{mz} obtain $\fe\la 0.2\solar$ from 1T
(Raymond-Smith) fits to the PSPC spectra of NGC 2563, HCG 62, and NGC
5846 within radii of $\sim 20\arcmin$ (though $\fe\sim 0.6\solar$ is
obtained for NGC 533). The values for NGC 2563 and 5846 obtained by
Mulchaey \& Zabludoff are inconsistent with those obtained from the
very similar analysis by \citet{hp} and from our deprojection
analysis.

A possible explanation could be associated with the background
subtraction. As we noted in \S \ref{caveats} the Fe abundances will be
underestimated if the background is also underestimated. This effect
is only noticeable when the background is underestimated by factors or
$\sim 2$ or more. Such a large factor will result if the background
spectrum is taken from the edge of the field and not corrected for
vignetting before subtracting from a source region near the center of
the field. As explained in \S \ref{bkg} we always account for the
different detector responses for the source and background in our
analysis.

However, when ignoring the response differences we are able to
reproduce the very small Fe abundances obtained by Mulchaey \&
Zabludoff for NGC 2563, 5846, and HCG 62. (We perform the 1T fits in
the same manner as described above for our comparison to the results
of Helsdon \& Ponman; i.e. we include all of the annuli investigated.)
Our value for NGC 533 is consistent within the large statistical error
with that obtained by Mulchaey \& Zabludoff whether or not we properly
scale the background to the source positions.  Since the Fe abundances
that we have determined from the PSPC data agree with all other
previous \rosat studies except Mulchaey \& Zabludoff, we are lead to
conclude that they must not have corrected their background spectra
for vignetting before subtracting from the source spectra.

Hence, there is no evidence from the \rosat PSPC data that the
brightest elliptical galaxies and galaxy groups have average Fe
abundances that are very sub-solar -- regardless of whether they have
steep Fe abundance gradients. The only clear evidence for very
sub-solar Fe abundances is thus derived from fits to the \asca data
assuming an isothermal gas which we have shown to give misleading
results because of the Fe bias.  Thus, it is the assumption of
isothermality, rather than the presence of Fe abundance gradients,
which accounts for the very sub-solar Fe abundances inferred by most
previous \asca studies (see Buote 2000a for a detailed review).

\subsection{Low Fe Abundance as the Result of Over-Smoothing}
\label{fp}

\begin{figure*}[t]
\parbox{0.49\textwidth}{
\centerline{\psfig{figure=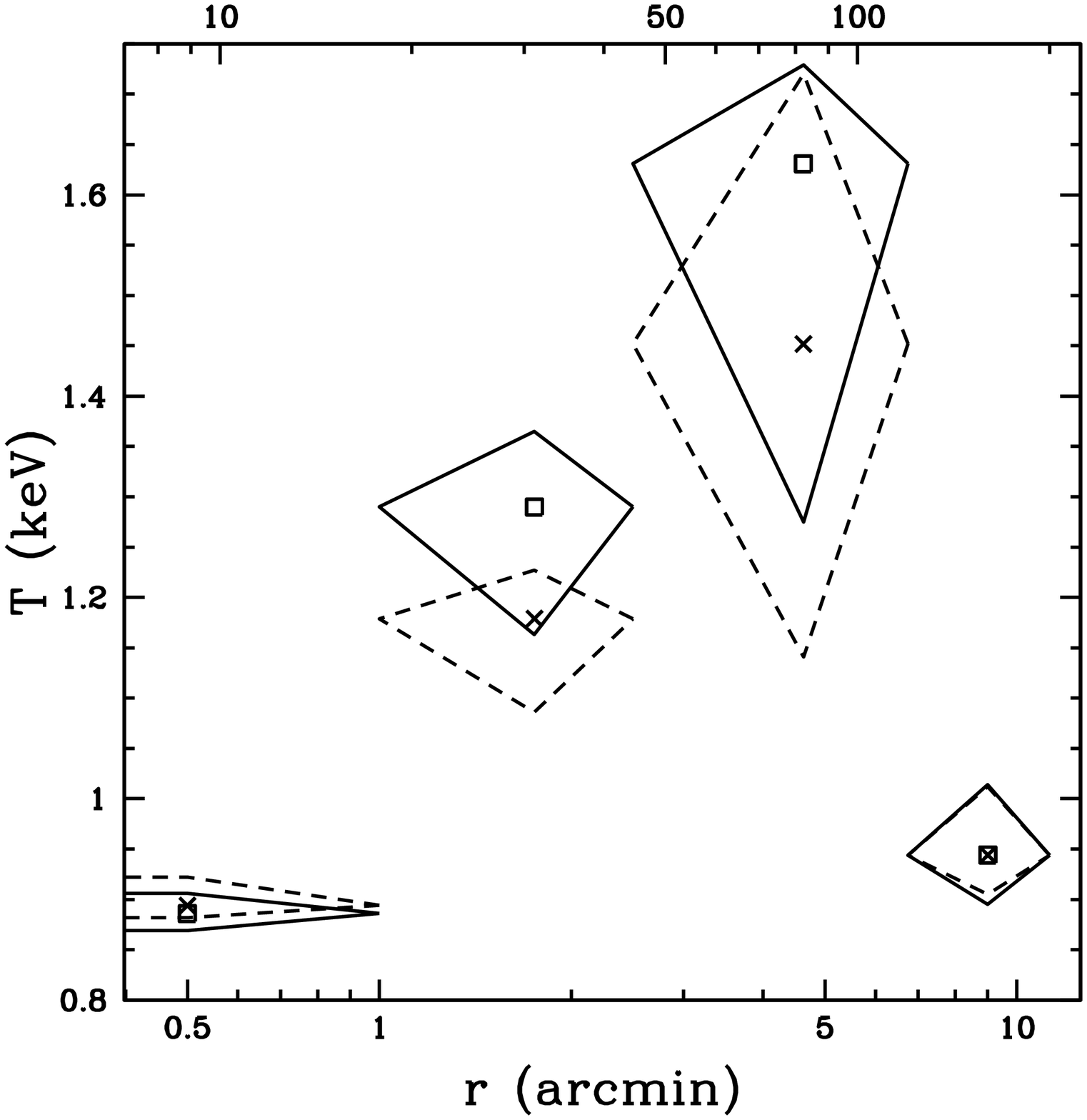,height=0.3\textheight}}
}
\parbox{0.49\textwidth}{
\centerline{\psfig{figure=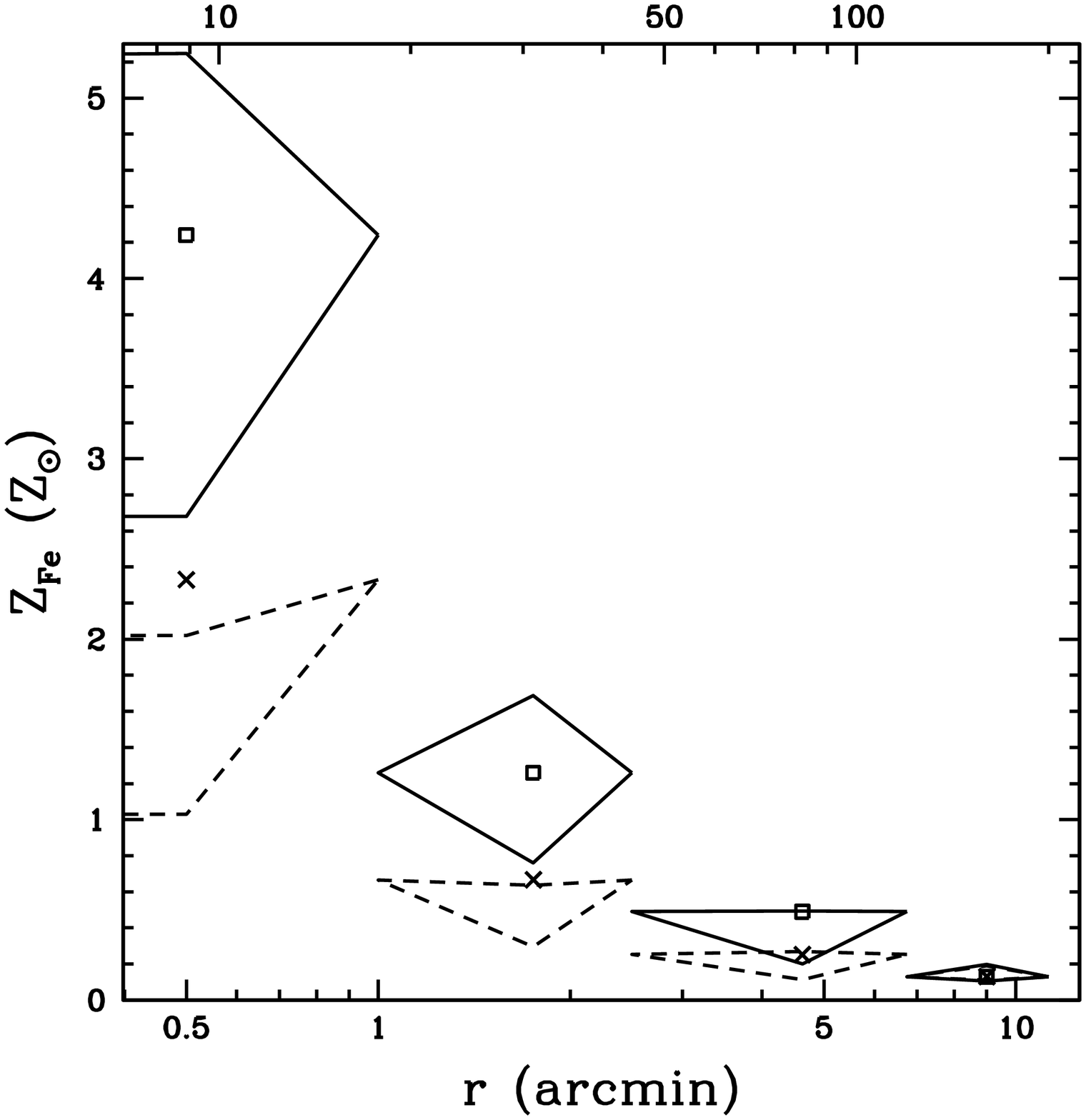,height=0.3\textheight}}
}
\caption{\label{fig.fp} Sensitivity to regularization criteria: (Left)
Temperature and (Right) Fe abundance profiles of HCG 62 obtained from
deprojection analysis of the \rosat data for models with fixed
Galactic column density (and no added edge). The squares and solid
diamonds represent models where the absolute values of the radial
logarithmic derivatives of the temperature and Fe abundance are
required to be $\le 1$ and $\le 2$ respectively which are the same
criteria we used for the models of HCG 62 in \S \ref{results}. The
crosses and dashed diamonds represent the case where the Fe abundance
derivative is required to be $\le 1$.  To facilitate the comparison we
have used the old photospheric solar Fe abundance (Fe/H is $4.68\times
10^{-5}$) as is done in \citet{fp} which leads to smaller values of
\fe\, than obtained in Figure \ref{fig.4ann}.}
\end{figure*}

The consistent picture that we have described is not supported by the
work of Finoguenov and co-workers who find very sub-solar Fe
abundances from deprojection analysis of the \asca data of HCG 62, NGC
4472, 4636, 4649, 5044, and 5846 (Finoguenov et al 1999; Finoguenov \&
Ponman 1999; Finoguenov \& Jones 1999). For HCG 62 and NGC 5044
\citet{fp} also perform a deprojection of the \rosat data and obtain
results not too dissimilar from their \asca analysis. As noted before
in \S \ref{results} we emphasize that our temperature profiles for
these systems agree well with those obtained in the Finoguenov papers.

We have previously speculated on possible reasons for the different
Fe abundances obtained by Finoguenov (see \S 4.1 of Buote 2000a). Now
equipped with our own deprojection code we can test one of those
speculations: the regularization assumptions. As discussed in \S
\ref{fluc} \citet{fp} require that the temperatures and abundances
vary approximately logarithmically with radius. However, we have shown
that most of these systems, especially HGC 62, have steep Fe abundance
gradients and thus these regularization criteria must be carefully
checked.

In Figure \ref{fig.fp} we plot the radial profiles of temperature and
Fe abundance of HCG 62 obtained from the \rosat deprojection analysis
where we have used the old photospheric solar Fe abundances for
consistent comparison with Finoguenov \& Ponman. Because of the steep
Fe abundance gradient we chose a slope of 2 to bound the radial
logarithmic Fe abundance derivative. (We arrived at this choice after
trying several different values and from comparison to results
obtained without deprojection -- see \S \ref{fluc}.) We also show the
results when the maximum Fe abundance derivative is set to 1. These
latter results agree well with the \asca and \rosat deprojection
results of Finoguenov \& Ponman; e.g., for $r\sim 30$ kpc we obtain
$\fe\sim 0.6\solar$ which is within the 90\% limits of their \asca
result and in excellent agreement with their \rosat value.  This value
is to be contrasted with $\fe\sim 1.3\solar$ obtained at that radius
when we set the maximum Fe abundance derivative to 2.

Therefore, we attribute the smaller Fe abundances obtained by
Finoguenov \& Ponman for HGC 62 to their overly restrictive
regularization criteria. Interestingly, unlike their \asca analysis of
HCG 62, their \rosat deprojection does not perform actual
regularization but rather adopts simple smooth analytical models which
apparently achieve the same effect. Finoguenov \& Ponman do apply
their regularization method to the deprojection of the \rosat data of
NGC 5044, and we can only reproduce their temperatures and small
Fe abundances if we require the radial logarithmic derivatives of both
the temperature and Fe abundance to be $\le 0.2$!  In fact, given the
steep Fe abundance gradients we obtain for most of the systems analyzed
in the Finoguenov papers over-smoothing probably accounts for most of
the Fe abundance differences between our papers.

\subsection{Implications and Future Work}

\begin{figure*}[t]
\centerline{\psfig{figure=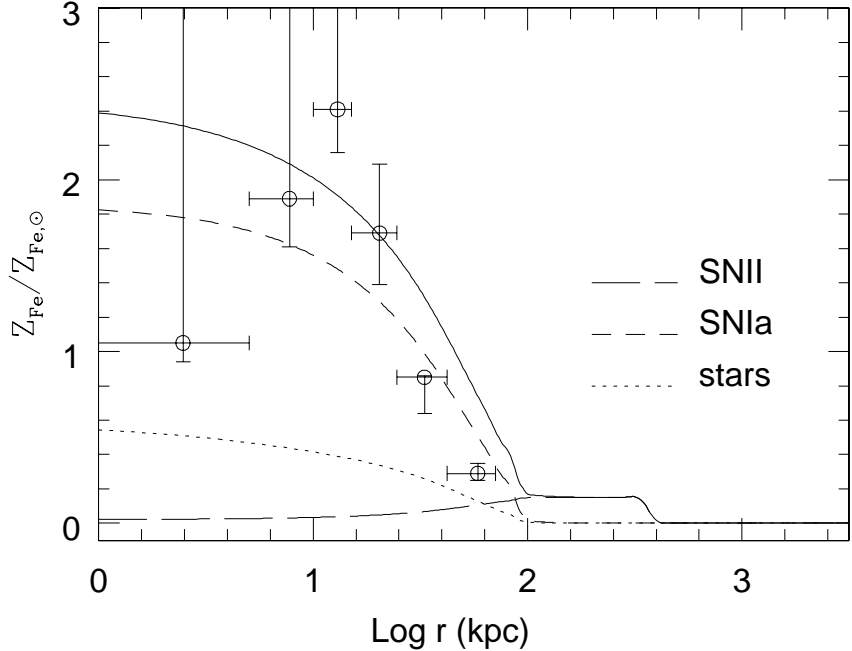,height=0.4\textheight}}
\caption{\label{fig.n4472} Comparison of the measured Fe abundances of
NGC 4472 with a gas-dynamical metal enrichment model (F. Brighenti and
W. Mathews, in preparation). The model is identical to those described
by \citet{bm99} having a Galactic IMF and accretion of primordial gas
except that the SNIa rate is normalized to the observed value in
nearby E/S0 galaxies reported by \citet{cap}.  The data points refer
to those in Figure \ref{fig.6ann} for NGC 4472 using our favored model
where the standard absorber has fixed Galactic column density and
where an intrinsic oxygen edge is included in the model.}
\end{figure*}

The Fe abundance gradients of these bright, nearby cooling flow
galaxies and groups reflect the history of star formation and the
dynamical evolution of the hot gas in these systems. In Figure
\ref{fig.n4472} we compare the Fe abundances we have measured for NGC
4472 with those predicted from a gas-dynamical metal enrichment model
which assumes (1) a Galactic IMF, (2) accretion of primordial gas
arising from secondary infall, and (3) a Type Ia supernova (SNIa) rate
normalized to the observed value in nearby E/S0 galaxies reported by
\citet{cap}. The qualitative agreement is clear, though the deviations
at larger radii indicate that the ram pressure distortions discussed
by \citet{is} compromise to some extent the assumptions of spherical
symmetry and hydrostatic equilibrium in the gas-dynamical model.

The relative contributions of the supernovae to the Fe abundance
within $\sim 50$ kpc are consistent with the ``solar supernova
proportion'' defined by \citet{rcdp}: i.e., enrichment models of the
Milky Way indicate that 1/4 of the Fe in the Sun arises from Type II
supernovae (SNII) and the remaining 3/4 arises from
SNIa. Consequently, the approximately solar Fe abundances we have
found for the brightest galaxies and groups within $r\sim 50$ kpc are
consistent with models of the chemical enrichment of the Milky way and
allow for the possibility of a universal IMF (e.g., Renzini et
al. 1993; Renzini 1997, 2000; Wyse 1997).  The problems associated
with the theoretical interpretation of the very sub-solar Fe abundances
inferred from previous studies of these systems (e.g., Renzini et
al. 1993; Arimoto et al. 1997; Renzini 1997) are eliminated by the
approximately solar Fe abundances we have inferred from both \asca and
\rosat data of bright galaxies and groups.

The sub-solar Fe abundances at large radii require additional
explanation, such as the dilution arising from the accretion of
primordial material included in the model in Figure \ref{fig.n4472}.
An alternative explanation for negative Fe abundance gradients as the
consequence of the sedimentation of heavier elements (Fabian \&
Pringle 1977; see Qin \& Wu 2000 for a recent discussion) seems less
plausible since the drift velocities are generally similar to or less
than the flow velocities in a cooling flow.

We mention that the Fe abundance at large radius ($\ga 100$ kpc)
predicted by the model for NGC 4472 (Figure \ref{fig.n4472}) arises
entirely from SNII ejecta. These SNII explosions accompanied the
initial burst of star formation which ended at a redshift before $\sim
2$. Hence, the Fe abundance measured at such large radii is a direct
fossil record of the ISM abundance at high redshift. For those systems
where we have measured Fe abundances at $r\ga 100$ kpc we find
$\fe\sim 0.5\solar$. If they are indeed the result of the initial SNII
enrichment these large Fe abundances are consistent with the arguments
that large quantities of the metals at high redshift reside in hot
gaseous halos (Pettini 1999, 2000) and that there was a prompt initial
enrichment of the early universe (Renzini 1997, 2000).

Finally, \chandra and \xmm will substantially clarify our
understanding of the metal enrichment and gas dynamics in cooling
flows. Their superior spatial and spectral resolution will enable
abundances to be measured on scales $r\la 1$ kpc which is critical
since the \rosat data suggest that the gradients may be largest at
small radii (\S \ref{results}). These new missions will also allow
accurate mapping of the abundances of several $\alpha$ elements (e.g.,
Si) which is of special importance because the $\alpha$/Fe ratios
provide the strongest constraints on the relative numbers of Type II
and Type Ia supernovae. The detailed constraints on the supernovae
feedback in these nearby cooling flows will be invaluable for
interpreting the star formation process in galaxies at high redshift
\citep{cgm}.

\acknowledgements

It is a pleasure to thank the referee, J. Irwin, for helpful comments,
F. Kelly for programming advice, and W. Mathews for permission to
include Figure \ref{fig.n4472} in this paper.  This research has made
use of (1) data obtained through the High Energy Astrophysics Science
Archive Research Center Online Service, provided by the NASA/Goddard
Space Flight Center, and (2) the NASA/IPAC Extragalactic Database
(NED) which is operated by the Jet Propulsion Laboratory, California
Institute of Technology, under contract with NASA.  Support for this
work was provided by NASA through Chandra Fellowship grant PF8-10001
awarded by the Chandra Science Center, which is operated by the
Smithsonian Astrophysical Observatory for NASA under contract
NAS8-39073.

\appendix

\section{Limiting Case for Edge Correction Factor}
\label{appendix}

The edge correction factor, $f(R_{i-1},R_i)$, given by equation A8 of
\citet{deproj_new} evaluates to zero in both the numerator and
denominator when $R_{i-1}=0$. However, the limit is well-behaved there,
and after applying L'H\^{o}pital's rule we obtain,
\begin{equation}
\lim_{\epsilon\to 0}f(\epsilon,R_i) = 
{ \left(R_{m-1}+R_m\right)R_{m-1}R_m \over R_i^3}
\left(-1 + {2\over \pi}
\left[{2R_i\over R_m} + \cos^{-1}\left({R_i\over R_m}\right) - 
{R_i\over R_m}\sqrt{1-{R_i^2\over R_m^2}}
\right]
\right),
\end{equation}
where $R_i$ is the radius of the circle in question and
$(R_{m-1},R_m)$ are the inner and outer radii of the bounding
annulus.


\begin{thebibliography}{}
\bibitem[Allen et al (2000)]{swa_agn} Allen, S. W., Di Matteo, T., \&
Fabian, A. C., 2000, \mnras, in press (astro-ph/9910188)
\bibitem[Anders \& Grevesse(1989)]{ag} Anders E., \& Grevesse N.,
1989, Geochimica et Cosmochimica Acta, 53, 197
\bibitem[Arabadjis \& Bregman(1999)]{ab} Arabadjis, J. S., \&
Bregman, J. N., 1999, \apj, 514, 607
\bibitem[Arimoto et al(1997)]{arimoto} Arimoto, N., Matsushita, K.,
Ishimaru, Y., Ohashi, T., Renzini, A., 1997, \apj, 477, 128
\bibitem[Arnaud(1988)]{kaa} Arnaud, K. A., 1988, in Cooling Flows in
Clusters of Galaxies, ed. A. C. Fabian, (Kluwer: Dordrecht), 31
\bibitem[Arnaud(1996)]{xspec} 
Arnaud K., 1996, in Jacoby G. and Barnes J., eds., Astronomical Data
Analysis Software and Systems V, ASP Conf. Series volume 101, p17
\bibitem[Baluci\'{n}ska-Church \& McCammon(1992)]{phabs}
Baluci\'{n}ska-Church, M., \& McCammon, D., 1992, \apj, 400, 699
\bibitem[Binney et al(1990)]{bdi} Binney, J. J., Davies, R. L., \&
Illingworth, G. D., 1990, \apj, 361, 78
\bibitem[Brighenti \& Mathews (1999)]{bm99}
Brighenti, F., \& Mathews, W. G., 1999, \apj, 515, 542
\bibitem[Buote(1999)]{b99} Buote, D. A., 1999, \mnras, 309, 695
\bibitem[Buote(2000a)]{b00a} Buote, D. A., 2000a, \mnras, 311, 176
\bibitem[Buote(2000b)]{b00b} Buote, D. A., 2000b, \apjl, 532, L113(PAPER1)
\bibitem[Buote(2000c)]{b00c} Buote, D. A., 2000c, \apj, submitted
(PAPER3) (astro-ph/0001330)
\bibitem[Buote \& Canizares(1994)]{bc94} Buote, D. A., \& Canizares,
C. R., 1994, \apj, 427, 86
\bibitem[Buote \& Fabian(1998)]{bf} Buote, D. A., \& Fabian,
A. C. 1998, \mnras, 296, 977 
\bibitem[Buote et al(1999)]{bcf} Buote, D. A., Canizares, C. R., \&
Fabian, A. C. 1999, \mnras, 310, 483
\bibitem[Cappellaro et al(1997)]{cap} Cappellaro E., Turatto M.,
Tsvetkov D. Yu., Bartunov O. S., Pollas C., Evans R., \& Hamuy M.,
1997, \aap, 322, 431
\bibitem[Cavaliere et al(2000)]{cgm} Cavaliere, A., Giacconi, R., \&
Menci, N., 2000, \apjl Letters, in press (astro-ph/9912374)
\bibitem[Chen et al(1997)]{cfg} Chen, L.-W., Fabian, A. C., \&
Gendreau, K. C., 1997, \mnras, 285, 449
\bibitem[Ciotti et al(1991)]{ciotti}
Ciotti L., D'Ercole A., Pellegrini S., \& Renzini A., 1991, \apj, 376,
380 
\bibitem[David et al(1991)]{david91}
David L. P., Jones C., \& Forman W., 1991, \apj, 380, 39
\bibitem[David et al (1994)]{djfd} David, L. P., Jones, C., Forman,
W., Daines, S., 1994, \apj, 428, 544
\bibitem[Dickey \& Lockman(1990)]{dl}
Dickey J. M., Lockman F. J., 1990, \araa, 28, 215
\bibitem[Fabian (1994)]{acf} 
Fabian A. C., 1994, \araa, 32, 277
\bibitem[Fabian \& Pringle(1977)]{fabprin} Fabian, A. C., \& Pringle,
J. E., 1977, \mnras, 181, 5p
\bibitem[Fabian et al(1981)]{deproj} Fabian, A. C., Hu, E. M., Cowie,
L. L., Grindlay, J., 1981, \apj, 248, 47
\bibitem[Feldman(1992)]{feld} Feldman U., 1992, Physica Scripta, 46,
202
\bibitem[Finoguenov \& Jones(2000)]{fj} Finoguenov, A., \& Jones, C.,
1999, \apj, submitted
\bibitem[Finoguenov \& Ponman(1999)]{fp} Finoguenov, A., \& Ponman,
T. J., 1999, \mnras, 305, 325
\bibitem[Finoguenov et al(1999)]{fino}
Finoguenov A., Jones C., Forman W., \& David L., 1999, \apj, 514, 844
\bibitem[Forman et al(1993)]{forman} Forman W., Jones C., David L.,
Franx M., Makishima K., \& Ohashi T., 1993, \apjl, 418, L55
\bibitem[Hasinger et al(1995)]{pspcpsf} Hasinger, G., Boese, G.,
Predehl, P., Turner, T. J., Yusaf, R., George, I. M., \& Rohrbach, G.,
1995 MPE/OGIP Calibration Memo CAL/ROS/93-015 ver. 1995 May 8
\bibitem[Helsdon \& Ponman(2000)]{hp} Helsdon, S. F., \& Ponman, T. J.,
2000, \mnras, in press (astro-ph/0002051)
\bibitem[Irwin \& Sarazin(1996)]{is} Irwin, J. A., Sarazin, C. L.,
1996, \apj, 471, 663
\bibitem[Ishimaru \& Arimoto(1997)]{im} Ishimaru, Y., \& Arimoto, N.,
1997, \pasj, 49, 1
\bibitem[Johnstone et al(1992)]{rjcool} Johnstone R. M., Fabian A. C.,
Edge A. C., \& Thomas P. A., 1992, \mnras, 255, 431
\bibitem[Jones et al(1997)]{j97} Jones, C., Stern, C., Forman, W.,
Breen, J., David, L., Tucker, W., \& Franx, M., 1997, \apj, 482, 143
\bibitem[Kaastra \& Mewe(1992)]{km}
Kaastra J. S., \& Mewe R., 1993, \aaps, 97, 443
\bibitem[Kim \& Fabbiano(1995)]{kf} Kim, D.-W., \& Fabbiano, G., 1995,
\apj, 441, 182
\bibitem[Kriss et al(1983)]{kcc} Kriss, G. A., Cioffi, D. F., \&
Canizares, C. R., 1983, \apj, 272, 439
\bibitem[Liedahl et. al(1995)]{mekal} 
Liedahl D. A., Osterheld A. L., \& Goldstein W. H., 1995, \apjl, 438,
L115 
\bibitem[Loewenstein \& Mathews(1991)]{lm}
Loewenstein M., Mathews W. G., 1991, \apj, 373, 445
\bibitem[Matsumoto et. al(1997)]{matsumoto} Matsumoto H., Koyama K.,
Awaki H., Tsuru T., Loewenstein M., \& Matsushita K., 1997, \apj, 482, 133
\bibitem[McLaughlin(1999)]{deproj_new} McLaughlin, D. E., 1999, \aj,
117, 2398
\bibitem[Mewe et al(1985)]{mewe} Mewe R., Gronenschild E. H. B. M., \&
van den Oord G. H. J., 1985, \aaps, 62, 197
\bibitem[Morrison \& McCammon(1983)]{wabs} Morrison, R., \& McCammon, D.,
1983, \apj, 270, 119
\bibitem[Mulchaey \& Zabludoff(1998)]{mz} 
Mulchaey J. S., \& Zabludoff A. I., 1998, \apj, 496, 73
\bibitem[Nulsen \& B\"{o}hringer(1995)]{nb} Nulsen, P. E. J., \&
B\"{o}hringer, H., 1995, \mnras, 274, 1093
\bibitem[Pettini(1999)]{max99} Pettini, M., 1999, in Chemical
Evolution from Zero to High Redshift, ed. J. Walsh and M. Rosa
(Berlin: Springer)(astro-ph/9902173) 
\bibitem[Pettini(2000)]{max00} Pettini, M., 2000, Philosophical
Transactions of The Royal Society: Series A (astro-ph/0001075)
\bibitem[Plucinsky et al(1993)]{pluc} Plucinsky, P. P., Snowden,
S. L., Briel, U. G., Hasinger, G., \& Pfeffermann, E., 1993, \apj,
418, 519
\bibitem[Ponman \& Bertram(1993)]{pb} Ponman, T. J., \& Bertram, D.,
1993, \nat, 363, 51
\bibitem[Press et al(1992)]{numrec} Press, W. H., Teukolsky, S. A.,
Vetterling, W. T., \& Flannery, B. P., 1992, Numerical Recipies in C
Second Edition (Cambridge Univ. Press: Cambridge)
\bibitem[Qin \& Wu(2000)]{qw} Qin, B., \& Wu, X.-P., 2000, \apjl, in
press (astro-ph/9912007) 
\bibitem[Rangarajan et al(1995)]{vijay} Rangarajan, F. V. N., Fabian,
A. C., Forman, W. R., \& Jones, C., 1995, \mnras, 272, 665
\bibitem[Renzini et al(1993)]{rcdp} Renzini A., Ciotti L., D'Ercole
A., \& Pellegrini S., 1993, \apj, 419, 52
\bibitem[Renzini(1997)]{r97} Renzini, A., 1997, \apj, 488, 35
\bibitem[Renzini(2000)]{r00} Renzini, A., 2000, in Large Scale
Structure in the X-Ray Universe, ed.  I. Georgantopoulos and
M. Plionis (Gyf sur Yvette, Ed. Frontieres) (astro-ph/0001312)
\bibitem[Sarazin(1986)]{sarazin} Sarazin, C. L., 1986,
Rev. Mod. Phys., 58, 1
\bibitem[Snowden(1999)]{quicksim} Snowden, S., 1999,
http://heasarc.gsfc.nasa.gov/docs/xmm/quicksim/
\bibitem[Snowden et al(1994)]{snow2} Snowden, S. L., McCammon, D.,
Burrows, D. N., \& Mendenhall, J. A., 1994, \apj, 424, 714
\bibitem[Trinchieri et al(1997)]{tfk}
Trinchieri G., Fabbiano G., Kim D.-W., 1997, A\&A, 318, 361
\bibitem[Trinchieri et al(1994)]{trin} Trinchieri G., Kim D.-W.,
Fabbiano G., \& Canizares C., 1994, \apj, 428, 555
\bibitem[Wyse(1997)]{wyse} Wyse, R. F. G., 1997, \apjl, 490, L69
\end{thebibliography}
\end{document}